\documentclass[aps,prd,twocolumn,superscriptaddress,groupedaddress,nofootinbib,showpacs,eqsecnum,useAms]{revtex4-1}

\usepackage{hyperref}
\usepackage{epsfig}
\usepackage{mathtools}
\usepackage{verbatim}
\usepackage{euscript,graphicx}
\usepackage{amsthm,dsfont,amsfonts,amsmath,amssymb}
\usepackage{euscript,color,fontenc,textcomp,relsize}
\usepackage{bm,url,float,cleveref}
\usepackage[caption=false]{subfig}
\allowdisplaybreaks

\begin{document}

\preprint{APS/123-QED}

\title{ Ready-to-use Fourier domain templates for compact binaries inspiraling along moderately eccentric orbits}
%Amplitude-corrected fourier domain gravitational waveforms for inspiralling eccentric binaries }% Force line breaks with \\
%\thanks{A footnote to the article title}%

\author{Srishti Tiwari\,$^{1}$ \footnote{srishti.tiwari@tifr.res.in}, Achamveedu Gopakumar\,$^{1}$, Maria Haney\,$^{2}$, and Phurailatapam Hemantakumar\,$^{3}$}%
\affiliation{
 $^{1}$Department of Astronomy and Astrophysics, Tata Institute of Fundamental Research, Mumbai 400005, India \\ $^{2}$Physik-Institut, Universit\"{a}t Z\"{u}rich, Winterthurerstrasse 190, 8057 Z\"{u}rich, Switzerland \\ $^{3}$Indian Institute of Technology, Mumbai 400076, India}
%\author{Srishti Tiwari}
%\email{srishti.tiwari@tifr.res.in}
%\affiliation{Department of Astronomy and Astrophysics, Tata Institute of Fundamental Research, Mumbai 400005, India}

%\author{Achamveedu Gopakumar} 
%\affiliation{Department of Astronomy and Astrophysics, Tata Institute of Fundamental Research, Mumbai 400005, India}

%\author{Maria Haney}
%\affiliation{Physik-Institut, Universit\"{a}t Z\"{u}rich, Winterthurerstrasse 190, 8057 Z\"{u}rich, Switzerland}

%\author{Phurailatapam Hemantakumar} 
%\affiliation{Indian Institute of Technology, Mumbai 400076, India}

\date{\today}
\begin{abstract}
We derive analytic expressions that provide Fourier domain gravitational wave (GW) response function for 
compact binaries inspiraling along moderately eccentric orbits. These expressions include
amplitude corrections to the two GW polarization states that are accurate to the 
first post-Newtonian (PN) order. Additionally, our fully 3PN accurate GW phase evolution
incorporates eccentricity effects up to sixth order at each PN order.
Further, we develop a prescription to incorporate analytically the effects of 3PN accurate periastron advance 
in the GW phase evolution.
This is how we provide a ready-to-use and efficient inspiral template family for 
compact binaries in moderately eccentric orbits.
Preliminary GW data analysis explorations suggest that our template family should be required to 
construct analytic inspiral-merger-ringdown templates to model moderately eccentric 
compact binary coalescence.

%\item[Usage]
%Secondary publications and information retrieval purposes.
%\item[PACS numbers]
%May be entered using the \verb+\pacs{#1}+ command.
%\item[Structure]
%You may use the \texttt{description} environment to structure your abstract;
%use the optional argument of the \verb+\item+ command to give the category of each %item. 
%\end{description}
\end{abstract}

% PACS, the Physics and Astronomy
                             % Classification Scheme.
%\keywords{Suggested keywords}%Use showkeys class option if keyword
                              %display desired
\maketitle
%\tableofcontents

\section{\label{sec:level1}Introduction}

 Observations of GW events by the advanced LIGO and VIRGO  GW interferometers 
 are ushering in the era of GW astronomy \cite{aLigo,aVirgo}.
%http://adsabs.harvard.edu/abs/2016PhRvL.116m1103A
% http://adsabs.harvard.edu/abs/2015CQGra..32b4001A
These GW events include merging black hole (BH) binaries and an inspiraling neutron star (NS) binary \cite{GW150914,GW151226,GW170104,GW170608,GW170814,GW170817,GWTC-1}.
%http://adsabs.harvard.edu/abs/2015CQGra..32b4001A
%http://adsabs.harvard.edu/abs/2016PhRvL.116f1102A
%http://adsabs.harvard.edu/abs/2016PhRvL.116x1103A
%http://adsabs.harvard.edu/abs/2017PhRvL.118v1101A
%http://adsabs.harvard.edu/abs/2017ApJ...851L..35A
%http://adsabs.harvard.edu/abs/2017PhRvL.119n1101A
%http://adsabs.harvard.edu/abs/2017PhRvL.119p1101A
Several scenarios 
%Many detailed and recent investigations suggest that several formation scenarios 
that include long-lived (galactic) field binaries, star clusters, galactic nuclei
and active galactic nuclei can produce these observed GW events \cite{Fieldbinaries,Globclusters,Galnuclei,AGN,LISAGCs}.
%Field binaries  https://arxiv.org/abs/1602.04531
%Cluster https://arxiv.org/abs/1703.01568
%G.nuclei https://arxiv.org/abs/1706.09896
%AGN https://arxiv.org/abs/1702.07818
Fortunately, it may be possible to extract valuable 
information about the astrophysical origins of GW events in the near future.
This requires accurate GW measurements of the spin-orbit misalignment or 
the orbital eccentricities of these GW events \cite{RZV16,CA17,NSBK17}.
%https://arxiv.org/abs/1609.05916
%https://arxiv.org/abs/1702.08479
%https://arxiv.org/abs/1606.09295
Using both frequency and time domain 
inspiral-merger-ringdown (IMR) waveforms, residual orbital eccentricities of 
the first two GW events were restricted to be below $0.15$ when these binaries entered aLIGO frequency window\cite{GW150914prop,Huerta}.
% http://adsabs.harvard.edu/abs/2016PhRvL.116x1102A
%https://arxiv.org/abs/1609.05933
Strictly speaking, the so far detected GW events do not exhibit any observational signatures of residual orbital eccentricities
and are faithfully captured by IMR templates associated with compact binaries merging along quasi-circular orbits.

 However, there exists a number of astrophysical scenarios that can 
produce GW events with non-negligible eccentricities in the frequency windows
of ground-based GW detectors. 
Dense star clusters like the ubiquitous globular clusters 
are the most promising sites to form aLIGO relevant compact binaries with non-negligible 
orbital eccentricities \cite{RCR16}.
% http://adsabs.harvard.edu/abs/2016PhRvD..93h4029R
A recent realistic modeling of globular clusters  that involve 
general relativistic few body interactions provided non-negligible 
fraction of BH binaries with eccentricities $> 0.1$ as they enter the aLIGO frequency window \cite{SMR14,SR17,RACR18,Samsing18,RACK18,LISAGCs}.
%http://adsabs.harvard.edu/abs/2018PhRvL.120o1101R
% http://adsabs.harvard.edu/abs/2018PhRvD..97j3014S
Additionally, there exists a number of other astrophysical scenarios that can force 
stellar mass compact binaries to merge with orbital eccentricities.
This include GW induced merger during hyperbolic encounters between 
BHs in dense clusters \cite{RYB16}
% http://iopscience.iop.org/article/10.3847/2041-8205/824/1/L12/meta
and mergers influenced by Kozai effect in few body systems as explored in 
many detailed investigations (see Ref.~\cite{Randall18} and references therein).
% https://arxiv.org/abs/1802.05718
Further, a very recent investigation pointed out that less frequent
binary-binary encounters in dense star clusters can easily produce 
eccentric compact binary coalescence \cite{ZSRHR18}.
% https://arxiv.org/abs/1810.00901
These detailed investigations suggest that it may be reasonable to expect GW events with non-negligible orbital eccentricities in the coming years.
Non-negligible orbital eccentricities may be helpful to improve 
the accuracy with a network of GW interferometers to 
constrain parameters of compact binary mergers \cite{GKRF18,Gondan_Kocsis}.
% https://arxiv.org/abs/1705.10781
Moreover, massive BH binaries in eccentric orbits are of definite 
interest to maturing Pulsar Timing Arrays and the planned 
Laser Interferometer Space Antenna (LISA) \cite{PTA_e,LISA_e}.
%https://ui.adsabs.harvard.edu/#abs/arXiv:1811.08826
%https://ui.adsabs.harvard.edu/#abs/2018arXiv181201011B/abstract

 There are different on-going investigations to model eccentric compact 
 binary coalescence.
These efforts aim to provide template families that model 
GWs from IMR phases of eccentric coalescence.
The initial effort, detailed in Ref.~\cite{Huerta}, 
%link.aps.org/pdf/10.1103/PhysRevD.95.024038
provided a time-domain IMR family that requires orbital eccentricity to be negligible during 
the merger phase. 
The inspiral part of the above waveform family was based on certain $x$-model, introduced in Ref.~\cite{HHLSxmodel},
that adapted GW phasing formalism of Refs.~\cite{DGI,KG06}. Additionally, 
a preliminary comparison with two numerical relativity (NR) waveforms was also pursued in Ref.~\cite{Huerta}.
An improved version of the above family was presented in Ref.~\cite{HKP18} 
%https://arxiv.org/abs/1709.02007
that employed certain quasi-circular merger waveform and which can reproduce 
their NR simulations for any mass ratio below $4$.
These waveform families are expected to model GWs from eccentric coalescence 
when initial eccentricities were usually below $0.2$.
Very recently, another time domain IMR family was introduced in Ref.~\cite{ENIGMA}.
%https://arxiv.org/abs/1711.06276
This detailed effort 
 combined various 
elements from  post-Newtonian, self-force and black hole perturbation 
approaches in tandem with NR simulations to model GWs from  moderately eccentric non-spinning
BH binary coalescence.  
The resulting IMR waveforms were validated with many NR 
simulations for 
eccentric binary BH mergers lasting around ten orbits with mass ratios below $5.5$ 
and initial eccentricities below $0.2$.
The eccentric binary BH coalescence is also explored in the framework of 
the Effective-One-Body (EOB) approach \cite{TD_review}.
%http://adsabs.harvard.edu/abs/2016LNP...905..273D
A formalism to incorporate orbital eccentricity in the existing EOB approach
to model quasi-circular compact binary coalescence is presented in Ref.~\cite{HB17}.
%arXiv:1707.08426
Additionally, Ref.~\cite{CHEOBNR} presented an EOB waveform family that incorporated elements of 2PN accurate
eccentric orbital description while comparing with few NR simulations for eccentric binary BH coalescence.
% arXiv:1708.00166
In contrast, the LIGO Scientific Collaboration (LSC) adapted 
Ref.~\cite{EMLP13} 
%https://arxiv.org/abs/1212.0837
 that provided a crude IMR prescription to model 
GW signals from merging highly eccentric compact binaries.
This was employed to probe the ability of few LSC algorithms to 
extract burst-like signals in the LIGO
data \cite{VTIWARI16}.
%https://arxiv.org/abs/1511.09240
Further, some of us developed a
ready-to-use {`effective eccentric variant'} of  \texttt{IMRPhenomD} waveform to
constrain 
the initial orbital eccentricity  of the GW150914 black hole binary.
This was pursued to justify the assumption of binary evolution
 along circular orbits  for the event \cite{GW150914prop}.
 A crucial ingredient of the above IMR waveform family involved 
  an eccentric version of 
\texttt{TaylorF2} approximant that incorporated in its
 Fourier phase the  
leading-order eccentricity corrections up to 3PN order.
The present paper provides fully analytic frequency domain interferometric response function $\tilde{h}(f)$ relevant for GW data analysis of nonspinning compact binaries inspiraling along moderately eccentric PN-accurate orbits.

Our computation is aimed at extending the widely used \texttt{TaylorF2} approximant that provides analytic 
 frequency domain GW templates for compact binaries inspiraling along quasi-circular orbits \cite{TaylorF2}.
This waveform family employs the method of stationary phase approximation (SPA)
%detailed in Refs.~ \cite{TaylorF2}, 
to compute analytically Fourier transform of temporally evolving GW polarization states, $h_{\times}$ and $h_{+}$, for quasi-circular inspirals.
The popular LSC approximant provides fully analytic Fourier domain 
GW response function $\tilde{h}(f)$ that incorporates  
3.5PN-accurate Fourier phase \cite{TaylorF2}.In other words, this approximant provides general relativistic corrections to GW phase evolution that are accurate to $( v/c)^7$ order beyond the  dominant quadrupolar order, where $v$ is the orbital velocity.
The present manuscript details our derivation of a fully analytic $\tilde{h}(f)$ with 3PN-accurate Fourier phase with  sixth order
eccentricity contributions in terms of certain initial eccentricity at each PN order. Additionally, we include 1PN-accurate amplitude corrections and the effect of 3PN-accurate periastron advance on the Fourier phases.
  
 To derive our eccentric approximant, we extend the post-circular scheme of Ref.~\cite{YABW} to higher PN orders.
This scheme involves expanding the Newtonian accurate $h_{\times}$ and $h_{+}$ as a power series in 
orbital eccentricity that requires analytic solution to the classic Kepler equation.
We extend such a Newtonian approach by invoking a recent effort
to solve analytically PN-accurate Kepler equation in the small eccentricity limit \cite{YS}. This detailed computation also provided analytic 1PN-accurate amplitude corrected expressions for $h_{\times}$ and $h_{+}$ as a sum over harmonics in certain mean anomaly $l$ of PN-accurate Keplerian type parametric 
solution \cite{YS}. Additionally, the above PN-accurate decomposition explicitly 
incorporated the effect of periastron advance on individual harmonics, numerically explored using PN description in Ref.~\cite{MG07}. We combine such 1PN-accurate amplitude corrected $h_{\times}$ and $h_{+}$ 
expressions that incorporated  eccentricity contributions to sixth order
at each PN order with the two beam pattern functions, $F_{\times}$ and $F_{+}$, to obtain fully analytic time domain GW response function $h(t)$.
Our eccentric \texttt{TaylorF2} approximant is obtained by applying the method of stationary phase approximation to such an analytic 
$h(t)=F_+ h_++F_\times h_\times$ expression.

To obtain analytic expressions for several Fourier phases at their associated stationary points of $h(t)$, we require 
additional PN-accurate expressions.
This involves deriving 3PN-accurate expression for the time eccentricity $e_t$, present in the 3PN-accurate Kepler Equation \cite{MGS}, as a bivariate expansion in terms of orbital angular frequency $\omega$, its initial value $\omega_0$ and $ e_0$, the value of $e_t$ at $\omega_0$. This lengthy computation extends to 
3PN order, the idea of certain 
{\it asymptotic eccentricity invariant} at the quadrupolar order, introduced in Ref.~\cite{KKS95}, and extended to 2PN in Ref.~\cite{THG}.
In fact, we adapted the approach of Ref.~\cite{THG} by employing the appropriately modified 3PN-accurate  $d\omega/dt$ and $de_t/dt$ expressions of Refs.~\cite{ABIS,KBGJ} 
to obtain 3PN-accurate bivariate expression for $e_t$. 
A careful synthesis of the above listed PN-accurate expressions lead to a fully analytic frequency domain \texttt{TaylorF2} approximant
that included 1PN-accurate amplitude corrections and 3PN-accurate Fourier phases. An additional feature of our approximant 
is the inclusion of periastron advance effects to 3PN order.
To explore GW data analysis implications of these features, we perform preliminary {\it match} computations \cite{DIS98}.
We conclude that the influences of periastron advance are 
non-negligible for moderately eccentric binaries, especially 
in the aLIGO frequency window.
This observation should be relevant while constructing IMR waveform family for compact binaries merging along moderate eccentric orbits.

This paper is structured as follows. In Sec.~\ref{sec:PostCirc}, we summarize the efforts of Refs.~\cite{YABW,THG} 
to obtain analytic $\tilde{h}(f)$ with PN-accurate Fourier phase. The crucial inputs to construct our eccentric \texttt{TaylorF2}
approximant is also listed in this section.
Our approach and crucial expressions to implement our eccentric approximant that incorporates eccentricity contributions up to $\mathcal{O}(e_t^6)$ to 3PN are presented in Sec.~\ref{sec:level2}.
A brief summary and possible extensions are listed in Sec.~\ref{conclusion} while 
detailed expressions, accurate to ${\cal O}(e_0^4)$ are given in Appendix \ref{appendixA}

\section{  Post-circular extensions to circular  inspiral templates } \label{sec:PostCirc}
 We begin by reviewing two key efforts to include the effects of orbital eccentricity onto the circular inspiral templates \cite{KKS95,YABW}.
This involves listing in Sec.~\ref{sec:PostCirc_1}  the steps that are crucial to compute 
 analytic frequency domain GW response function with quadrupolar amplitudes and PN-accurate Fourier phase 
 in some detail. 
 Various lengthy expressions, extracted from Refs.~\cite{YS,ABIS,KBGJ}, are listed in Sec.~\ref{sec:PostCirc_2}
 %http://adsabs.harvard.edu/abs/2018PhRvD..98j4043K
 that will be 
 crucial to compute the time domain response function for eccentric binaries
 while incorporating effects of periastron advance, higher order radiation reaction and amplitude corrections.

 \subsection{\label{sec:PostCirc_1} Quadrupolar order $\tilde{h}(f)$ with PN-accurate Fourier phase } 
 
 Following \cite{KT89_300}, we may express 
the GW interferometric response function as
\begin{widetext}
\begin{equation}
h(t)= F_+\left(\theta_S,\phi_S,\psi_S\right)  h_+(t) +  F_\times\left(\theta_S,\phi_S,\psi_S\right) h_\times (t) \,,  \label{ht_ant} \end{equation}
\end{widetext}
where $F_{\times,+}\left(\theta_S,\phi_S,\psi_S\right)$ 
are the two detector antenna patterns.
These quantities depend on $\phi_S,\theta_S$, the right ascension
and declination of the source, and certain  polarization angle 
$\psi_S$ \cite{KT89_300}. 
For eccentric inspirals, the explicit expressions for the quadrupolar order GW polarization states, $h_{\times}$ and $h_+$,  are 
given by Eqs.~(3.1) of Ref.~\cite{YABW}.
%while using true anomaly or by Eqs.~() of Ref.~\cite{}
%while using a combination of  eccentric and true anomalies.
It is rather straightforward to express these Newtonian accurate expressions as a sum over harmonics 
in terms of the mean anomaly $l$.
%, detailed in Ref.~ \cite{YABW}.
The resulting expressions read
\begin{widetext}
\begin{equation}       
h_{+,\times}(t)=    -   \frac{G m \eta}{c^2 D_L}   \, x\,   
   \sum\limits_{j=1}^{10}    \left[   C_{+,\times}^{(j)} \cos{ j l}  + S_{+,\times}^{(j)} \sin{j l}    \right]\,,   
\label{Eq_1}        
\end{equation}
\end{widetext}
where $D_L$ denotes the luminosity distance while the symmetric mass ratio $\eta$ of a binary consisting of individual masses 
$m_1$ and $m_2$ is defined to be $\eta= (m_1\,m_2)/ m^2$ while the total mass  
$m= m_1 + m_2$.
Further, we use the commonly used dimensionless PN expansion parameter 
 $x=\,\left(\frac{G\, m\, \omega}{c^3}\right)^{2/3}$ where $G$, $c$ and $\omega$ are the
 gravitational constant, the speed of light in vacuum and the orbital angular frequency,
 respectively.
 % NOTE:DEFINITION OF x PARAMETER ADDED}
% Also $G$ and $c$ stand for the gravitational constant and the speed of light respectively.
The  Newtonian accurate amplitudes, 
 $C^{(j)}_{+,\times}$ and $S^{(j)}_{+,\times}$, are written as  
power series in orbital eccentricity $e_t$ whose coefficients involve  
  trigonometric functions of the two angles 
$\iota, \beta$ that specify the line of sight vector in a certain
inertial frame.
The derivation of these expressions is detailed in Ref.~\cite{YABW} and the required
inputs are obtained by adapting a standard analytic approach to solve the classical Kepler equation 
in terms of the Bessel functions\cite{PC_93}.
% http://adsabs.harvard.edu/abs/1993sket.book.....C

 With the help of  Eqs.~(\ref{ht_ant}) and (\ref{Eq_1}), we obtain interferometric strain for GWs from  eccentric binaries as 
 \begin{equation} 
 \label{Eq_hN}
h(t) = \, -   \frac{G m \eta}{c^2 D_L}   \left(\frac{G m \omega}{c^3} \right)^{2/3}   \sum\limits_{j=1}^{10}  \alpha_j~\cos\left(j l +\phi_j\right) ,     
%\label{Eq_2}     
\end{equation} 
where $\alpha_j =  {\rm sign}(\Gamma_j) \sqrt{\Gamma_j^{2}+\Sigma_j^{2}}  $ and $ \phi_j = \tan^{-1}{\left(-\frac{\Sigma_j}{\Gamma_j}\right)} $. 
The two new functions, 
$ \Gamma_j$ and $ \Sigma_j$,  are defined as
$\Gamma_j=
 F_+\, C^{(j)}_+ + F_{\times}\, C^{(j)}_{\times}$
and $\Sigma_j =   F_+\, S^{(j)}_+ + F_{\times}\, S^{(j)}_{\times}  $, respectively as in Ref.~\cite{YABW}.
We impose the effects of GW emission on the above strain by 
specifying how $e_t$ and $\omega = 2\, \pi\, F$, 
$F$ being the orbital frequency,  vary in time.
 In Ref.~\cite{YABW}, the temporal evolutions of $\omega$ and $e_t$ are 
governed by the following Newtonian (or quadrupolar) equations
that were adapted from Refs.~\cite{PM63,Peters64,JS92}.
\begin{widetext}
\begin{subequations}         \label{4}
\begin{align}           
\frac{d \omega}{dt} &=
\frac { (G\,m\ \omega)^{5/3}\,\omega^{2}\,\eta}
{5\,c^5\, (1 -e_t^2)^{7/2}}
\biggl \{  96
+292\,{{e_t}}^{2}
+37\,{{e_t}}^{4} \biggr \}          ,           \label{4.b}\\    
\frac{d e_t}{dt} &=
-\frac { (G\,m\, \omega)^{5/3}\, \omega \,\eta \, e_t }
{ 15\,c^5\, (1 -e_t^2)^{5/2}}
\biggl \{ 304 +  121\,{{e_t}}^{2} \biggr \} \,.        \label{4.c}
\end{align}
\end{subequations}
\end{widetext}
%{\bf See Eqs at the end of the document}
It is customary to solve these two coupled differential equations 
numerically to obtain $\omega(t)$ and $e_t(t)$ and hence 
temporally evolving $h(t)$. 
 Interestingly, earlier efforts provided certain analytic 
 way for obtaining temporal evolution for $\omega(t)$ and $e_t(t)$ 
that mainly involves the usage of hypergeometric functions \cite{MRLY18,MKFV,Pierro02,Pierro01}

%which can make it computationally expensive 

However, it is possible to  obtain analytic frequency domain counterpart of the above $h(t)$ as demonstrated in Ref.~\cite{KKS95,YABW}.
This traditional approach involves the  method of
SPA, detailed in Ref.~\cite{BO_book}, to compute  analytically
the Fourier Transform of $h(t)$.
This was essentially demonstrated at the leading order
in initial eccentricity $e_0$ in Ref.~\cite{KKS95} and later extended to
${\cal O}(e_0^8)$ in Ref.~\cite{YABW}.
Following Refs.~\cite{KKS95,YABW}, we write 
\begin{widetext}
\begin{align}
\tilde{h}(f) =  \mathcal{\tilde{A}} \,   {\left(\frac{G m \pi f}{c^3}\right)}^{-7/6}     \sum\limits_{j=1}^{10} \xi_{j}
{\left(\frac{j}{2}\right)}^{2/3}  e^{-i(\pi/4 + \Psi_j)}\,,           \label{5} 
\end{align}
\end{widetext}
where the overall amplitude  $\mathcal{\tilde{A}}$ and 
the amplitudes of Fourier coefficients $ \xi_j $ 
%that incorporate ${\cal O}(e_t^8)$ eccentricity contributions 
are given by 
\begin{subequations}                         \label{6}  
\begin{align}
 \mathcal{\tilde{A}} &= - {\left(\frac{5 \eta \pi}{384}\right)}^{1/2}  \frac{G^2 m^2}{c^5 D_L}  ,   \label{6.b}\\              
 \xi_j  &=  \frac{\left(1-e_t^2\right)^{7/4}}{{\left(1+\frac{73}{24}e_t^2+\frac{37}{96}e_t^4\right)}^{1/2}} \alpha_{j} e^{-i \phi_j(f/j) }    .                                      \label{6.c}                                
\end{align}
\end{subequations}
In the approach of stationary phase approximation, the crucial Fourier phase 
is given by
\begin{align}
\Psi_j [F(t_0)] = 2 \pi  \int_{}^{F(t_0)}     \tau' \left( j - \frac{f}{F'}   \right)    \,d{F'}\,,               \label{7} 
\end{align}
where $\tau$ stands for $F / \dot {F}$.
Note that one needs to evaluate 
the above integrals  at appropriate  
stationary points $t_0$, defined by $F(t_0) = f/j$.

 To obtain a fully analytic ready-to-use expression for $\tilde{h}(f)$, we need to follow few additional steps.
Clearly, we require to specify the frequency evolution of $e_t$  with the help of above Eqs.~(\ref{4}).
%(\ref{4.b}) and (\ref{4.c}).
The structure of these equations for $\dot \omega$ and $\dot e_t$ allows us to write 
$ d \omega/d e_t = \omega\, \kappa_N (e_t) $
and it turns out that $\kappa_N $ depends only on $e_t$.
This allows to integrate analytically the resulting 
$ d \omega / \omega = \kappa_N (e_t)\, d e_t$ equation.
The resulting expression can be written symbolically as 
$ \omega  / \omega_0 = \kappa' (e_t, e_0)$ where $e_0$ is the 
value of $e_t$ at the initial $\omega$ value, namely $\omega_0$
(see Eq.~(62) in Ref.~\cite{DGI} for the explicit form for 
$\kappa' (e_t, e_0)$).
Interestingly, one may invert such an expression in the limit 
 $e_t \ll 1$ to obtain 
 $e_t$ in terms of $e_0, \omega$ and $\omega_0$ and 
it reads  
\begin{align}
e_t &\sim e_0 \chi ^{-19/18}  + \mathcal{O}(e_0^3)  ,         \label{7.c}
\end{align}
where $\chi$ is defined as $ \omega/ \omega_0 = F/F_0$.
We note that the above result was first obtained 
in Ref.~\cite{KKS95} which influenced them to introduce the idea
of an {\it asymptotic eccentric invariant }.
%The above result was extended to
%${\cal O}(e_0^8)$ in Ref.~\cite{YABW}.
This relation allows us to write 
$\tau $ in terms of 
 $\omega, \omega_0$ and $e_0$ as
\begin{widetext}
\begin{align}
\tau  \sim    \frac{5~}{96~\eta~x^4}\left(\frac{G~m}{c^3}\right) \left[ 1-\frac{157 e_0^2}{24 }\chi ^{-19/9} + \mathcal{O}(e_0^4)\right]  .            \label{7.d}
\end{align}
\end{widetext}
%%%%\tau &\sim \frac{5 M}{384\ 2^{2/3} (F M)^{8/3} \pi ^{8/3}} \left[1-\frac{157 e_0^2}{24 \chi ^{19/9}} + {\cal O}(e_0^4)
It is now straightforward to compute analytically the indefinite integral
for $\Psi_j$, namely 
\begin{align}
2 \pi  \int     \tau'\left( j - \frac{f}{F'}   \right)    \,d{F'} 
\end{align}
 that appears 
in Eq.~(\ref{7}) for $\tilde{h}(f)$.
This leads to the following 
 expression for $\Psi_j$,  accurate to $ {\cal O}(e_0^2)$ corrections: 
\begin{widetext}
\begin{align}
 \Psi_j &\sim j \phi_c - 2\pi f t_c 
   - \frac{3}{128 \eta     }{  \left( \frac{G m \pi f }{c^3}    \right)   }^{-5/3} \left(\frac{j}{2}\right)^{8/3}
   \left[1-\frac{2355 e_0^2}{1462 }\chi ^{-19/9} + \mathcal{O}(e_0^4)     \right ]\,,              \label{8}
\end{align}
\end{widetext}
where $\phi_c$ and $t_c$ are the orbital phase at coalescence and the time of coalescence, respectively.
Note that $\chi$  now stands for $f/f_0$
due to the use of the stationary phase condition. 
%the $\chi$ in the above equation stands for $f/f_0$.
Additionally, we have re-scaled $F_0 \rightarrow f_0/j$
to ensure
that $e_t(f_0) = e_0$ while employing the above expression for $e_t$, given by Eq.~(\ref{7.c}).
Indeed, our expression is consistent with Eq.~(4.28) of Ref.~\cite{YABW} that employs the 
chirp mass to characterize the binary.
A number of extensions to the above result is available in the literature.
In fact, Ref.~\cite{YABW} computed the 
higher order corrections to $e_t$
in terms of $e_0$  up to
${\cal O}( {e_0^7})$
 and  extended  $\Psi_j$ to ${\cal O}( {e_0^8})$.
 Its PN extension,  available in
 Ref.~\cite{THG}, provided 
2PN corrections for $\Psi_j$ that incorporated
eccentricity corrections, accurate to 
  ${\cal O}( {e_0^6})$ at every PN
order while Ref.~\cite{MFAM16} computed 3PN-accurate $\Psi_j$  
that included leading order $e_0$ contributions.
 
 A crucial ingredient to such PN extensions is the derivation of PN-accurate 
 $e_t$ expression in terms of $e_0, \chi$ and $x$.
 % . NOTE: DEFINITION OF x MOVED BELOW EQ. 2.2. LINES EDITED HERE}
 In what follows, we summarize the steps that are required to obtain 
1PN-accurate expression for $e_t$ (see Ref.~\cite{THG} for details).
The starting point of such a derivation is the 1PN-accurate differential equations 
for $\omega $ and $e_t$, obtainable from Eqs.~(3.12) in Ref.~\cite{THG}.
With these inputs, it is fairly straightforward to obtain the following 1PN accurate expression
for $ d \omega/ \omega $ 
that includes only the leading order $e_t$ contributions as 
\begin{widetext}
\begin{align}
 d \omega/ \omega
 =  \biggl \{ -\frac{18}{19 e_t} -\frac{3}{10108 e_t} \left(-2833+5516 \eta\right) \left(\frac{G m \omega}{c^3}\right)^{2/3} 
\biggr \}\, de_t 
  \,. \label{12}
\end{align}
\end{widetext}
The fact that $\omega$ term appears only at the 1PN order 
allows us to use the earlier derived Newtonian accurate 
$\omega = \omega_0\, \left ( e_0/e_t \right )^{18/19}$ relation to
replace $\omega$ on the right hand side of the above equation.      
This leads to
%leads to the following 1PN-accurate equation for $ d \omega/ \omega$:
\begin{widetext}
\begin{align}
 d \omega/ \omega \sim \left\lbrace-\frac{18}{19 e_t} -  \frac{3}{10108}   \left(\frac{e_0^{12/19}}{e_t^{31/19}}\right)
    \left(-2833+5516\eta\right)\, x_0 
    %\left(\frac{G m \omega_0}{c^3}\right)^{2/3}
  \right\rbrace de_t\,,      \label{13}
\end{align}
\end{widetext}
where  $ x_0= \left ( G\, m\, \omega_0/c^3 \right )^{2/3}$. We can integrate this equation to obtain $ \ln \omega - \ln \omega_0$ in terms of $e_t,e_0$ and $\omega_0$.
The exponential of the resulting expression and its
bivariate expansion in terms of $x_0$ and $e_t$ result in 
\begin{widetext}
\begin{align}
\omega & \sim \left\lbrace  \left(\frac{e_0}{e_t}\right)^{18/19} + x_0\left(\frac{2833-5516 }{2128}\eta   \right)     \left[  \left( \frac{e_0}{e_t}\right)^{18/19}     -       \left( \frac{e_0}{e_t}\right)^{30/19}      \right]             \right\rbrace   \omega_0  \,.
  \label{14}
\end{align}
\end{widetext}
 We invert the above equation to obtain $e_t$ in terms of $e_0$ and $x_0$ after invoking the 
Newtonian accurate relation $e_t = e_0\, \chi^{-19/18} $ to 
 replace the $e_t$ terms associated with the $x_0$ term.
This inversion and the associated bivariate 
expansion in terms of $e_0$ and $x_0$ require that $e_0 \ll 1$ and $x_0 \ll 1$.
The resulting $e_t$ expression reads
\begin{widetext}
\begin{align}
&e_t \sim e_0 
\left\lbrace  \chi ^{-19/18}+x_0      \left(\frac{2833}{2016}-\frac{197 }{72}\eta\right)    \left(- \chi ^{-7/18}   +  \chi
   ^{-19/18}   \right)    
\right\rbrace \,.                   \label{15}
\end{align}
\end{widetext}
To obtain $e_t$ as a bivariate expansion in terms of
the regular PN parameter $x$ and $e_0$, we employ the fact that $ x/x_0 = \chi^{2/3}$ and this results in 
% that incorporates the leading 
%order $e_0$ corrections both at the Newtonian and 1PN order
%  reads
\begin{widetext}
\begin{align}
 &e_t \sim e_0 \left\lbrace   \chi^{-19/18}   +  x\left(   \frac{2833}{2016}-\frac{197 }{72}\eta    \right)\left( -\chi^{-19/18} +\chi^{-31/18} \right)              \right\rbrace         .   \label{16}
\end{align}
\end{widetext}
We are now in a position to obtain
1PN-accurate $\Psi_j$ expression that includes 
 ${\cal O}(e_0^2)$ contributions both at the Newtonian and 1PN orders
with the help of 1PN-accurate $\tau = \omega/\dot \omega$ 
expression that is accurate to  ${\cal O}(e_t^2)$ terms.
A straightforward computation leads to the desired  $\Psi_j$ expression which reads
 \begin{widetext}
\begin{align}
 &\Psi_j \sim     j \phi_c- 2\pi f t_c  -   \left(\frac{3 j  }{256 \eta } \right) x^{-5/2}     \left\lbrace 1- \frac{2355 e_0^2 }{1462} \chi^{-19/9} +  x \left[ \frac{3715}{756}                 +\frac{55  
   }{9}\eta +   \left(   \left[   -\frac{2045665  }{348096}- \frac{128365 }{12432}\eta\right] \chi^{-19/9}   \right.\right.\right.\nonumber\\       
  &\qquad \left.\left.\left. {}      +  \left[ -\frac{2223905}{491232 }+\frac{154645  }{17544 }\eta\right]\chi^{-25/9}        \right)e_0^2 \right]\right\rbrace     \,,            \label{19}
 \end{align}
 \end{widetext}
where the quantities $x$ and $\chi$ will have to be evaluated 
at the stationary point (see Ref.~\cite{THG} for details). 
With the above equation, we explicitly listed our approach to compute PN-accurate 
$ \Psi_j$ that incorporates $e_0$ corrections at each PN order. 
In the present paper, we extend these computations to 3PN order 
while incorporating ${\cal O}(e_0^6)$ contributions at each PN order.
These higher order $e_0$ corrections are included as we desire to to model
GWs from moderately eccentric compact binary inspirals.
In the next section, we provide crucial inputs that will be required to 
compute analytic  1PN-accurate amplitude corrected $\tilde{h}(f)$ with 3PN-accurate Fourier phases.

\subsection{Analytic PN-accurate amplitude corrected time domain eccentric GW templates} \label{sec:PostCirc_2}

 The previous section showed that we require analytic expressions for the two GW polarization states 
as a sum over {\it harmonics} to construct ready-to-use analytic $\tilde{h}(f)$.
This influenced us to adapt Eqs.~(44) and (45) in Ref.~\cite{YS} that 
provided analytic 1PN-accurate amplitude corrected  $h_{\times,+}(t)$ 
which additionally included the 
effects of periastron advance on individual {\it harmonics}.
This may be seen by a close inspection of appropriate terms in Eqs.~(44),(45),(46) and (47) of Ref.~\cite{YS}.
%The use of Eqs.~(44) and (45) in Ref.~\cite{YS} to construct GW response function does affect the structure  of $h(t)$.
To describe in detail how these improvements
in GW polarization states 
change the harmonic structure of $h(t)$, we restrict 
our attention to quadrupolar order  contributions to $h_{\times,+}(t)$ , given in Eqs.~(44) and (45) of Ref.~\cite{YS}.
The explicit expressions for such `Newtonian` contributions to $h_{\times,+}(t)$ that include $\mathcal{O}(e_t^4)$ corrections read
\begin{widetext}
\begin{align}
h_{\times}^{0}=& \,\frac{G\,m\,\eta}{c^2\, D_L} \,x\, \bigg\{\cos(\phi+\phi')\bigg[\bigg(-3 e_t+\frac{13 e_t^3}{8}\bigg) c_i s_{2 \beta }\bigg] +\sin(\phi+\phi')\bigg[\bigg(3 e_t-\frac{13 e_t^3}{8}\bigg) c_i c_{2 \beta }\bigg] +\cos(2\phi)\bigg[\bigg(4-10 e_t^2 \nonumber \\ & +\frac{23 e_t^4}{4}\bigg) c_i s_{2 \beta }\bigg]+ \sin(2\phi)\bigg[\bigg(-4+10 e_t^2-\frac{23 e_t^4}{4}\bigg) c_i c_{2 \beta }\bigg]+\cos(3\phi-\phi')\bigg[\bigg(9 e_t-\frac{171 e_t^3}{8}\bigg) c_i s_{2 \beta }\bigg] \nonumber \\ & +\sin(3\phi-\phi')\bigg[\bigg(-9 e_t+\frac{171 e_t^3}{8}\bigg) c_i c_{2 \beta }\bigg]+\cos(4\phi-2\phi')\bigg[\bigg(16 e_t^2-40 e_t^4\bigg) c_i s_{2 \beta }\bigg]+\sin(4\phi-2\phi')\bigg[\bigg(-16 e_t^2 \nonumber \\ & +40 e_t^4\bigg) c_i c_{2 \beta }\bigg]+\cos(5\phi-3\phi')\bigg[\frac{625}{24} e_t^3 c_i s_{2 \beta }\bigg]+\sin(5\phi-3\phi')\bigg[\frac{-625}{24}  e_t^3 c_i c_{2 \beta }\bigg]+\cos(6\phi-4\phi')\bigg[\frac{81}{2} e_t^4 c_i s_{2 \beta }\bigg] \nonumber \\ & +\sin(6\phi-4\phi')\bigg[\frac{-81}{2} e_t^4 c_i c_{2 \beta }\bigg]+\cos(\phi-3\phi')\bigg[\frac{-7}{24} e_t^3 c_i s_{2 \beta }\bigg]+\sin(\phi-3\phi')\bigg[\frac{-7}{24} e_t^3 c_i c_{2 \beta }\bigg] \nonumber \\ & +\cos(2\phi-4\phi')\bigg[-\frac{1}{4} e_t^4 c_i s_{2 \beta }\bigg]+\sin(2\phi-4\phi')\bigg[-\frac{1}{4} e_t^4 c_i c_{2 \beta }\bigg] \bigg\} \label{hx0} \,, \\ \nonumber \\
h_+^{0}=& \,  \frac{G\,m\,\eta}{c^2\, D_L} \,x\,\bigg\{\cos(\phi+\phi')\bigg[\bigg(\frac{3 e_t}{2}-\frac{13 e_t^3}{16}\bigg) \left(1+c_i^2\right) c_{2 \beta }\bigg]+\sin(\phi+\phi')\bigg[\bigg(\frac{3 e_t}{2}-\frac{13 e_t^3}{16}\bigg) \left(1+c_i^2\right) s_{2 \beta }\bigg] \nonumber \\ & +\cos(2\phi)\bigg[\bigg(-2+5 e_t^2-\frac{23 e_t^4}{8}\bigg) \left(1+c_i^2\right) c_{2 \beta }\bigg]+\sin(2\phi)\bigg[\bigg(-2+5 e_t^2-\frac{23 e_t^4}{8}\bigg) \left(1+c_i^2\right) s_{2 \beta }\bigg] \nonumber \\ & +\cos(3\phi-\phi')\bigg[\bigg(-\frac{9 e_t}{2}+\frac{171 e_t^3}{16}\bigg) \left(1+c_i^2\right) c_{2 \beta }\bigg]+\sin(3\phi-\phi')\bigg[\bigg(-\frac{9 e_t}{2}+\frac{171 e_t^3}{16}\bigg) \left(1+c_i^2\right) s_{2 \beta }\bigg] \nonumber \\ & +\cos(4\phi-2\phi')\bigg[\left(-8 e_t^2+20 e_t^4\right) \left(1+c_i^2\right) c_{2 \beta }\bigg]+\sin(4\phi-2\phi')\bigg[\left(-8 e_t^2+20 e_t^4\right) \left(1+c_i^2\right) s_{2 \beta }\bigg] \nonumber \\ & +\cos(5\phi-3\phi')\bigg[-\frac{625}{48} e_t^3 \left(1+c_i^2\right) c_{2 \beta }\bigg]+\sin(5\phi-3\phi')\bigg[-\frac{625}{48} e_t^3 \left(1+c_i^2\right) s_{2 \beta }\bigg]+\cos(6\phi-4\phi')\bigg[-\frac{81}{4} e_t^4 \left(1+c_i^2\right) c_{2 \beta }\bigg] \nonumber \\ & +\sin(6\phi-4\phi')\bigg[-\frac{81}{4} e_t^4 \left(1+c_i^2\right) s_{2 \beta }\bigg]+\cos(\phi-\phi')\bigg[\bigg(e_t-\frac{e_t^3}{8}\bigg) s_i^2\bigg]+\cos(2\phi-2\phi')\bigg[\bigg(e_t^2-\frac{e_t^4}{3}\bigg) s_i^2\bigg] \nonumber \\ & +\cos(3\phi-3\phi')\bigg[\frac{9}{8} e_t^3 s_i^2\bigg]+\cos(4\phi-4\phi')\bigg[\frac{4}{3} e_t^4 s_i^2\bigg]+\cos(\phi-3\phi')\bigg[\frac{7}{48} e_t^3 \left(1+c_i^2\right) c_{2 \beta }\bigg] \nonumber \\ & +\sin(\phi-3\phi')\bigg[-\frac{7}{48} e_t^3 \left(1+c_i^2\right) s_{2 \beta }\bigg]+\cos(2\phi-4\phi')\bigg[-\frac{1}{8} e_t^4 \left(1+c_i^2\right) c_{2 \beta }\bigg]+\sin(2\phi-4\phi')\bigg[-\frac{1}{8} e_t^4 \left(1+c_i^2\right) s_{2 \beta }\bigg]\bigg\}\,. \label{hp0}
\end{align}
\end{widetext}
where $\phi= ( 1+k)\,l $, $ \phi' = k\,l$ and 
$k$ provides the rate of periastron advance per orbit \cite{DGI}.
Further, we let $c_i=\cos \iota$, $s_i=\sin \iota$, $c_{2\beta}=\cos 2\beta$ and $s_{2\beta}=\sin 2\beta$.
%with $\iota$ and $\beta$ as defined in Sec.\ref{sec:PostCirc_1}.
Note that crucial ingredients to obtain above analytic expressions 
include developing approaches to solve PN-accurate Kepler equation and 
adapting them to derive PN-accurate relations to connect true 
and eccentric anomalies, detailed in Ref.~\cite{YS}.
A close inspection of the above two equations with Eqs.~(3.3) and (3.4) of Ref.~\cite{YABW} reveals that the arguments of cosine and sine functions 
in above expressions involve $\phi'= k\,l$ and its multiples
in addition to the usual orbital phase $\phi$ and its multiples.
These additional $\phi'$ contributions 
%to the `usual' orbital phase variable $\phi$
are clearly due to the periastron advance.
It turns out that these additional angular contributions are 
sufficient to provide the numerically inferred side bands in the power spectrum of eccentric binaries due to the presence of $k$ \cite{MG07}.
 This is why we explicitly included $e_t^4$ contributions to the above $h_{\times,+}$
expressions  as these contributions are required to reveal 
the underlying side band structure of waveforms due to the influence of periastron advance.

 We re-write the above expressions for $h_{\times,+}^{0}$ in a more compact form to explicitly show how various harmonics are affected by the 
 advance of periastron. The resulting expressions read 
 \begin{widetext}
\begin{align}
h_{+,\times}^0(t)=& \,\bigg\{\sum_{j=1}^{6}\bigg[C_{+,\times}^{j,-2}(0) \,\cos(j\,\phi-(j-2)\phi') + S_{+,\times}^{j,-2}(0) \,\sin(j\,\phi-(j-2)\phi')\bigg] + \sum_{j=1}^4 \bigg[ C_{+,\times}^{j,0}(0) \,\cos(j\,\phi-j\phi')\nonumber \\ & + S_{+,\times}^{j,0}(0) \,\sin(j\,\phi-j\phi')\bigg]+ \sum_{j=1}^{2} \bigg[C_{+,\times}^{j,+2}(0) \,\cos(j\,\phi-(j+2)\phi') + S_{+,\times}^{j,+2}(0) \,\sin(j\,\phi-(j+2)\phi')\bigg]\bigg\} \,, \label{hpx0}
\end{align}
\end{widetext}
where we denoted the coefficient of $\cos (j\,\phi-(j\pm n)\phi')$ harmonic at the quadrupolar (Newtonian) order 
for the $+$ polarization by $C_{+}^{j,\pm n}(0)$ 
while the coefficient of $\sin (j\,\phi-(j\pm n)\phi')$ is indicated by 
$S_{+}^{j,\pm n}(0)$.
We adopt a rather heavy notation as it is amenable to higher PN order contributions which will be tackled below.
In this convention, we represent 
the coefficient of $\cos(j\,\phi-(j\pm n)\phi')$ that appears in the
 1PN contributions to  $\times $ polarization state by 
$C_{\times}^{j,\pm n}(1)$. It should be obvious that $j$ stands for the {\it harmonic} variable while $n$ provides a measure of the shift that each harmonic experiences due 
to periastron advance.
A close comparison of Eqs.~(\ref{hx0}) and (\ref{hp0}) reveals that these coefficients are functions of $\iota, \beta$ 
and contain powers of $e_t$.
Moreover, the arguments of cosine and sine functions clearly show that the eccentricity induced higher {\it harmonics} are not mere multiples of $\omega=N(1+k)$, where $N$ is the PN-accurate mean motion. Clearly, this is due to the presence of non-vanishing $\phi'$ contributions due to periastron advance.
 %that enters the arguments of these cosine and sine functions through  $\phi'$ angle.
Interestingly, the plus polarization state does provide {\it harmonics} which are
integer multiples of $N$.
It is not difficult to show that these Newtonian like terms arise from 
specific cosine functions with arguments $j\phi-j\phi' $, as evident 
from Eqs.~(\ref{hp0}).
%In our Eqs.~(\ref{hp0},\ref{hpx0}), these contributions arise from {\color{blue}{$j\phi-j\phi' $ terms}} which leads to $\cos j\,l$ contributions to %$h_+^{0}$.
Further, it is possible to show that these contributions arise from $e_t\, \cos u\,s_i^2/( 1- e_t\,\cos u)$ 
contributions to $H_{+}^{0}$, given by Eq.~(F2a) in Ref.~\cite{YS} and therefore not influenced by the periastron advance.
Interestingly,  similar 
%advance. Similar 
conclusions were obtained in Ref.~\cite{MG07}.

 With the above inputs, we write the time-domain GW detector response function for eccentric inspirals as
 \begin{widetext}
\begin{align}
h(t)=& \, \frac{G\,m\,\eta}{c^2\, D_L} \,x\, \bigg\{ \sum_{j=1}^6 \bigg[ \Gamma_{j,-2}^{(0)} \cos(j\phi-(j-2)\phi')+\Sigma_{j,-2}^{(0)} \sin(j\phi-(j-2)\phi') \bigg] + \sum_{j=1}^4 \bigg[ \Gamma_{j,0}^{(0)} \cos(j\phi-j\phi')\nonumber \\ & +\Sigma_{j,0}^{(0)} \sin(j\phi-j\phi') \bigg] + \sum_{j=1}^2 \bigg[ \Gamma_{j,+2}^{(0)} \cos(j\phi-(j+2)\phi')+\Sigma_{j,+2}^{(0)} \sin(j\phi-(j+2)\phi') \bigg] \bigg\}\,, \label{hpx0_1}
\end{align}
\end{widetext}
where the amplitudes of the cosine and sine functions are denoted by rather complicated symbols
$\Gamma_{j,\pm n}^{(0)} $ and  $\Sigma_{j,\pm n}^{(0)}$. The definition of 
$h(t)=\,F_+\,h_+(t)\,+\,F_\times\,h_\times(t)$ ensures that 
$\Gamma_{j,\pm n}^{(0)}=F_+\,C_{+}^{j,\pm n}(0) + F_\times\,C_{\times}^{j,\pm n}(0)$
while $\Sigma_{j,\pm n}^{(0)}=F_+\,S_{+}^{j,\pm n}(0) + F_\times\,S_{\times}^{j,\pm n}(0)$. We list in Appendix~\ref{appendixB}, the 
lengthy expressions for these quantities in terms of $\iota, \beta$ and eccentricity contributions, accurate to ${\cal O}(e_t^4)$.
% NOTE: LINES MODIFIED}
  We display up to $\mathcal{O}(e_t^4)$ contributions to demonstrate the full harmonic structure of the quadrupolar order GW polarization states.
It turns out that  $\Sigma_{j,0}^{(0)}$ contributions 
are zero by construction. This is mainly because 
the un-shifted harmonics only appear with the cosine terms, present in the $+$ polarization state.
%COMMENT: These terms have order e_t^4 contributions not e_t^6. So first sentence is incorrect. $\Sigma_{j,0}^{(0)}$ contributions are ALWAYS vanishing. The unshifted harmonic terms ONLY appear in PLUS polarization with COSINE terms. The reason being the presence of \cos u term in H_+ (See Yannick and Abhimanyu's paper) 
 Invoking familiar trigonometric identities, we simplify the above equation and obtain
%By employing the trigonometric identity for $\cos(a+b)=\cos(a)\cos(b)-\sin(a)\sin(b)$, we can further simplify Eq.(9) which then becomes
\begin{widetext}
 \begin{align}
\label{Eq_h_t_0f} 
h(t)=& \,\frac{G\,m\,\eta}{c^2\, D_L} \,x\,\bigg\{ \sum_{j=1}^6 \alpha_{j,-2}^{(0)} \cos(j\phi-(j-2)\phi'+\bar{\phi}_{j,-2}^{(0)}) + \sum_{j=1}^4 \alpha_{j,0}^{(0)} \cos(j\phi-j\phi'+\bar{\phi}_{j,0}^{(0)}) \nonumber \\ & + \sum_{j=1}^2 \alpha_{j,+2}^{(0)} \cos(j\phi-(j+2)\phi'+\bar{\phi}_{j,+2}^{(0)})\bigg\} \,,
\end{align}
\end{widetext}
where we introduce two new multi-index symbols $\alpha_{j,\pm n}^{(0)}$ and $\bar{\phi}_{j,\pm n}^{(0)}$ to ensure that 
detector strain can be written in terms of only cosine functions.
Influenced by Ref.~\cite{YABW}, these symbols are defined as 
$\alpha_{j,\pm n}^{(0)} = {\rm sign}\left(\Gamma_{j,\pm n}^{(0)}\right)\sqrt[]{\left(\Gamma_{j,\pm n}^{(0)}\right)^2+\left(\Sigma_{j,\pm n}^{(0)}\right)^2}$
and $\bar{\phi}_{j,\pm n}^{(0)} = \tan^{-1}\left(- \frac{\Sigma_{j,\pm n}^{(0)}}{\Gamma_{j,\pm n}^{(0)}}\right)$.
We do not list explicit expressions for these quantities that are accurate to $\mathcal{O}(e_t^4)$ in eccentricity corrections
as they can be easily obtained from our Eqs.~(\ref{gamma_0}) and (\ref{sigma_0}).

 A close inspection of above equations reveal that they provide 
GW response function for compact binaries moving along precessing eccentric orbits.
To obtain temporally evolving $h(t)$ associated with 
compact binaries inspiraling along precessing eccentric orbits,
we need to specify how $\phi, \phi', \omega $ and $e_t$ vary in time due to GW emission. We adapt the phasing formalism, detailed 
in Refs.~\cite{DGI,THG}, to provide differential equations for these
variables. And, for the time being, we will concentrate on the secular evolution of these 
variables. In other words, we will neglect GW induced  quasi-periodic variations to orbital elements and angles, detailed in Ref.~\cite{DGI}. The 3PN-accurate secular evolution to 
$\phi$ and $\phi'$ in the modified harmonic gauge 
that are 
%We employ the following 3PN-accurate differential equations that are 
accurate to ${\cal O}(e_t^6)$ are given by 
%in modified harmonic gauge :
\begin{widetext}
\begin{align} 
\frac{d \phi}{dt} =&\, \omega =\, x^{3/2}\frac{c^3}{G\,m}, \, \label{Tevolve_1} \\ \nonumber \\
\frac{d \phi'}{dt} =&\, \omega \frac{k}{1+k} =\, \omega \bigg\{3\,x \bigg[1+ e_t^2+ e_t^4+e_t^6\bigg]+x^2\bigg[\frac{9}{2}-7 \eta +\bigg(\frac{87}{4}-\frac{41}{2}\eta\bigg) e_t^2+(39-34 \eta ) e_t^4+\bigg(\frac{225}{4}-\frac{95}{2}\eta\bigg) e_t^6\bigg] \nonumber \\ & +x^3\bigg[\frac{27}{2}+\bigg(-\frac{481}{4}+\frac{123 \pi ^2}{32}\bigg) \eta +7 \eta ^2+\bigg(\frac{519}{4}+\bigg(-\frac{2037}{4}+\frac{1599 \pi ^2}{128}\bigg)\eta +61 \eta ^2\bigg) e_t^2+\bigg(\frac{2811}{8}\nonumber \\ & +\bigg(-1174+\frac{3321 \pi ^2}{128}\bigg) \eta +\frac{1361}{8}\eta^2\bigg)e_t^4+\bigg(\frac{10779}{16}+\bigg(-\frac{16901}{8}+\frac{2829 \pi ^2}{64}\bigg) \eta +\frac{2675}{8}\eta^2\bigg) e_t^6\bigg]\bigg\},\, \label{Tevolve_2} \\ \nonumber \\
\frac{d \omega}{dt} =&\, \frac{96\,c^6\,\eta}{5\,G^2\,m^2}x^{11/2}\bigg\{1+\frac{157}{24}e_t^2+\frac{605}{32}e_t^4+\frac{3815}{96}e_t^6+x\bigg[-\frac{743}{336}-\frac{11}{4}\eta+\bigg(\frac{713}{112}-\frac{673}{16}\eta\bigg) e_t^2+\bigg(\frac{52333}{672}-\frac{12415}{64}\eta\bigg)e_t^4 \nonumber \\ & +\bigg(\frac{13823}{48}-\frac{107765}{192}\eta\bigg) e_t^6\bigg]+ \dot{\omega}^{\tiny{1.5PN}} + \dot{\omega}^{\tiny{2PN}} + \dot{\omega}^{\tiny{2.5PN}} + \dot{\omega}^{\tiny{3PN}} \bigg\},
\, \label{Tevolve_3} \\ \nonumber \\
\frac{d e_t}{dt} =& \, -\frac{304\,c^3\,\eta\,e_t}{15\,G\,m}x^4\bigg\{1+\frac{881}{304}e_t^2+\frac{3265}{608}e_t^4+\frac{20195}{2432}e_t^6+x\bigg[-\frac{2817}{2128}-\frac{1021}{228}\eta+\bigg(\frac{40115}{4256}-\frac{51847}{1824}\eta\bigg) e_t^2 \nonumber \\ & +\bigg(\frac{87749}{2128} -\frac{298115}{3648}\eta\bigg) e_t^4+\bigg(\frac{121833}{1216}-\frac{2501905}{14592}\eta\bigg) e_t^6\bigg]+ \dot{e_t}^{\tiny{1.5PN}} + \dot{e_t}^{\tiny{2PN}} + \dot{e_t}^{\tiny{2.5PN}} + \dot{e_t}^{\tiny{3PN}}\bigg\}.
\label{Tevolve_4} \\ \nonumber 
\end{align}  
\end{widetext}

The explicit $1.5,\,2,\,2.5$ and $3$PN order contributions to  $d \omega/dt$ and $ d e_t/dt$
that incorporates all the 
 $\mathcal{O}(e_t^6)$ corrections are provided in the Appendix~\ref{appendixC}.
 %. NOTE: LINES MODIFIED} 
The differential equations for $d \omega/dt$ and $ d e_t/dt$ are extracted from expressions, available in Refs.~\cite{ABIS,KBGJ} and 
are in the modified harmonic gauge. These papers provided above 3PN accurate expressions as 
sum of certain `instantaneous' and `tail' contributions 
\begin{align*}
\frac{d \omega}{dt} &= \bigg( \frac{d \omega}{dt}\bigg)_{\rm inst} + \bigg( \frac{d \omega}{dt} \bigg)_{\rm tail}
\,,\\
\frac{d e_t}{dt} &= \bigg( \frac{d e_t}{dt} \bigg)_{\rm inst} + \bigg( \frac{d e_t}{dt}\bigg)_{\rm tail}\,.
\end{align*} 
The 3PN-accurate instantaneous contributions that depend only on the binary dynamics at the usual retarded time while the hereditary contributions  are sensitive to the binary dynamics at all epochs prior to the usual retarded time \cite{BDI}. The instantaneous contributions to $d \omega/dt$ are extracted from Eqs.~(6.14),(6.15a),(6.15),(C6) and (C7) of  Ref.~\cite{ABIS} while for $d e_t/dt$ such contributions originate from Eqs.~(6.16),(6.19a),(6.19b),(C10) and (C11) in Ref.~\cite{ABIS}. It should be obvious that we have Taylor expanded  these equations around $e_t=0$ to obtain eccentricity contributions
accurate to ${\cal O}(e_t^6)$. The hereditary contributions to $d \omega/dt$ and $d e_t/dt$ are adapted from Eqs.~(6.24c) and (6.26) 
of Ref.~\cite{ABIS} and they depend on a number of eccentricity enhancement functions.
We employ such enhancement functions provided in Ref.~\cite{KBGJ} for our computations.
%{\bf 
%The equations for $d \omega/dt$ and $de_t/dt$ are extracted from Ref.~\cite{ABIS}.
%Further, they are consistent with Eqs.~() in Ref.~\cite{}.
%The equation for $d \phi/dt$ arises by expressing $N \times k$ in terms of $\omega = N ( 1 +k)$.
%We list here the following fully 3PN-accurate expressions for $n$ and $k$ in terms of $\omega$ and $e_t$
%that are crucial to obtain Eq.~(). The relevant equations are
%\begin{widetext}
\newline
%We need to write few lines from our earlier draft.
%}
We now have all the inputs to 
obtain the restricted time-domain $h(t)$
to model GWs from non-spinning compact binaries inspiraling along precessing
moderately eccentric orbits.
To obtain such time domain templates, 
 we numerically solve 
the above listed differential equations for $\omega, e_t, \phi$ and $\phi'$
and impose their temporal evolution in the quadrupolar order GW response 
function, given by Eq.~(\ref{Eq_h_t_0f}).
We now move onto describe how we extend the quadrupolar order GW response 
function.
 
It should be obvious that we require a prescription to compute 
analytically PN-accurate amplitude corrected GW polarization states 
to improve the above listed quadrupolar order GW response 
function.
Therefore, we adapt 1PN-accurate amplitude corrected and fully analytic expressions for  $h_{\times,+}$, available in Ref.~\cite{YS},
to compute GW response function for eccentric inspirals that incorporates  PN contributions  even to its amplitudes.
We list below certain ingredients that will be crucial to write down analytic $h(t)$ that incorporates 
1PN-accurate amplitude corrections to $h_{\times,+}$ while consistently keeping eccentricity contributions up to
${\cal O}(e_t^6)$.
We begin by displaying Eqs.~(44) and (45) of Ref.~\cite{YS} as a single sum which reads 
\begin{widetext}
\begin{align} \label{hpc}
h_{+,\times}(t)=&\, \frac{G\,m\,\eta}{c^2\, D_L} \,x\, \bigg\{h^0_{+,\times}(t)+x^{0.5}\,h^{0.5}_{+,\times}(t)+x\,h^1_{+,\times}(t)\bigg\} \,.
\end{align} 
Various PN order amplitude contributions take the following form
\begin{subequations}
\begin{align}
h_{+,\times}^0(t)=& \,\sum_{j=1}^{8}\bigg\{ C_{+,\times}^{j,-2}(0) \,\cos(j\,\phi-(j-2)\phi') + S_{+,\times}^{j,-2}(0) \,\sin(j\,\phi-(j-2)\phi')\bigg\}+ \sum_{j=1}^6 \bigg\{ C_{+,\times}^{j,0}(0) \,\cos(j\,\phi-j\phi')\nonumber \\ & + S_{+,\times}^{j,0}(0) \,\sin(j\,\phi-j\phi') \bigg\} + \sum_{j=1}^{4} \bigg\{C_{+,\times}^{j,+2}(0) \,\cos(j\,\phi-(j+2)\phi') + S_{+,\times}^{j,+2}(0) \,\sin(j\,\phi-(j+2)\phi')\bigg\},  \label{22}\\
h_{+,\times}^{0.5}(t)=& \, \delta\bigg\{\sum_{j=1}^{7}\bigg[ C_{+,\times}^{j,-1}(0.5) \,\cos(j\,\phi-(j-1)\phi') + S_{+,\times}^{j,-1}(0.5) \,\sin(j\,\phi-(j-1)\phi')\bigg] + \sum_{j=1}^{5} \bigg[ C_{+,\times}^{j,+1}(0.5) \,\cos(j\,\phi-(j+1)\phi') \nonumber  \\ & + S_{+,\times}^{j,+1}(0.5) \,\sin(j\,\phi-(j+1)\phi')\bigg] + \sum_{j=1}^{9} \bigg[ C_{+,\times}^{j,-3}(0.5) \,\cos(j\,\phi-(j-3)\phi') + S_{+,\times}^{j,-3}(0.5) \,\sin(j\,\phi-(j-3)\phi') \bigg] \nonumber \\ & + \sum_{j=1}^{3} \bigg[ C_{+,\times}^{j,+3}(0.5) \,\cos(j\,\phi-(j+3)\phi') + S_{+,\times}^{j,+3}(0.5) \,\sin(j\,\phi-(j+3)\phi')\bigg]\bigg\}, \label{23} \\
h_{+,\times}^{1}(t)=& \, \sum_{j=1}^{8}\bigg\{C_{+,\times}^{j,-2}(1) \,\cos(j\,\phi-(j-2)\phi') + S_{+,\times}^{j,-2}(1) \,\sin(j\,\phi-(j-2)\phi')\bigg\}+ \sum_{j=1}^{4} \bigg\{C_{+,\times}^{j,+2}(1) \,\cos(j\,\phi-(j+2)\phi') \nonumber \\ & + S_{+,\times}^{j,+2}(1) \,\sin(j\,\phi-(j+2)\phi')\bigg\}+ \sum_{j=1}^{10} \bigg\{C_{+,\times}^{j,-4}(1) \,\cos(j\,\phi-(j-4)\phi') + S_{+,\times}^{j,-4}(1) \,\sin(j\,\phi-(j-4)\phi') \nonumber  \bigg\} \\ & + \sum_{j=1}^{2} \bigg\{C_{+,\times}^{j,+4}(1) \,\cos(j\,\phi-(j+4)\phi') + S_{+,\times}^{j,+4}(1) \,\sin(j\,\phi-(j+4)\phi') \bigg\}  + \sum_{j=1}^6 \bigg\{C_{+,\times}^{j,0}(1) \,\cos(j\,\phi-j\phi') \nonumber  \\ & + S_{+,\times}^{j,0}(1) \,\sin(j\,\phi-j\phi')\bigg\}, \label{24}
\end{align}
\end{subequations}
\end{widetext}
 where $\delta=(m_1-m_2)/(m_1+m_2)$ and we let
  $m_1$ to be the heavier of the two binary components. We do not list explicitly very lengthy expressions for these amplitudes. However, they can be easily extracted from the attached 
\texttt{Mathematica} notebook.
The derivation of above lengthy expressions include developing analytic approaches to solve PN-accurate Kepler equation and PN-accurate relations connecting true and eccentric anomalies, detailed in Ref.~\cite{YS}.
Indeed, we have verified that these expressions reduce to their circular 
counterparts, provided in Ref.~\cite{BIWW_95}.
%http://adsabs.harvard.edu/abs/1996CQGra..13..575B

The associated GW detector strain for eccentric binaries is given by 
\begin{widetext}
\begin{align}
h(t)=& \, \frac{G\,m\,\eta}{c^2\, D_L} \,x\, \bigg\{\bigg[ \sum_{j=1}^8 \bigg( \Gamma_{j,-2}^{(0)} \cos(j\phi-(j-2)\phi')+\Sigma_{j,-2}^{(0)} \sin(j\phi-(j-2)\phi') \bigg) + \sum_{j=1}^6 \bigg(\Gamma_{j,0}^{(0)} \cos(j\phi-j\phi')+\Sigma_{j,0}^{(0)} \sin(j\phi-j\phi') \bigg) \nonumber \\ &  + \sum_{j=1}^4 \bigg(\Gamma_{j,+2}^{(0)} \cos(j\phi-(j+2)\phi')+\Sigma_{j,+2}^{(0)} \sin(j\phi-(j+2)\phi') \bigg)\bigg] + x^{0.5}\,\delta \bigg[ \sum_{j=1}^7 \bigg(\Gamma_{j,-1}^{(0.5)} \cos(j\phi-(j-1)\phi') \nonumber \\ & +\Sigma_{j,-1}^{(0.5)} \sin(j\phi-(j-1)\phi') \bigg) + \sum_{j=1}^5 \bigg(\Gamma_{j,+1}^{(0.5)} \cos(j\phi-(j+1)\phi')+\Sigma_{j,+1}^{(0.5)} \sin(j\phi-(j+1)\phi') \bigg)\nonumber \\ & + \sum_{j=1}^9 \bigg(\Gamma_{j,-3}^{(0.5)} \cos(j\phi-(j-3)\phi')+\Sigma_{j,-3}^{(0.5)} \sin(j\phi-(j-3)\phi') \bigg) + \sum_{j=1}^3 \bigg(\Gamma_{j,+3}^{(0.5)} \cos(j\phi-(j+3)\phi') \nonumber \\ &  +\Sigma_{j,+3}^{(0.5)} \sin(j\phi-(j+3)\phi') \bigg)\bigg]+\,x \bigg[\sum_{j=1}^8 \bigg(\Gamma_{j,-2}^{(1)} \cos(j\phi-(j-2)\phi')+\Sigma_{j,-2}^{(1)} \sin(j\phi-(j-2)\phi') \bigg) \nonumber \\ & + \sum_{j=1}^4 \bigg(\Gamma_{j,+2}^{(1)} \cos(j\phi-(j+2)\phi')+\Sigma_{j,+2}^{(1)} \sin(j\phi-(j+2)\phi') \bigg) + \sum_{j=1}^6 \bigg(\Gamma_{j,0}^{(1)} \cos(j\phi-j\phi')+\Sigma_{j,0}^{(1)} \sin(j\phi-j\phi') \bigg) \nonumber \\ &  + \sum_{j=1}^{10} \bigg(\Gamma_{j,-4}^{(1)} \cos(j\phi-(j-4)\phi')+\Sigma_{j,-4}^{(1)} \sin(j\phi-(j-4)\phi') \bigg) + \sum_{j=1}^2 \bigg(\Gamma_{j,+4}^{(1)} \cos(j\phi-(j+4)\phi')\nonumber \\ & +\Sigma_{j,+4}^{(1)} \sin(j\phi-(j+4)\phi') \bigg)\bigg]\bigg\},
\end{align}
\end{widetext}
where as expected, we have defined 
\begin{subequations}
\begin{align}
\Gamma_{j,\pm n}^{(p)}&=F_+\,C_{+}^{j,\pm n}(p) + F_\times\,C_{\times}^{j,\pm n}(p),\\
\Sigma_{j,\pm n}^{(p)}&=F_+\,S_{+}^{j,\pm n}(p) + F_\times\,S_{\times}^{j,\pm n}(p).
\end{align}
\end{subequations}
A further simplification is possible which requires, as expected, additional multi-index functions 
\begin{subequations} \label{alpha_phibr}
\begin{align} 
\alpha_{j,\pm n}^{(p)}&=\,{\rm sign}\left(\Gamma_{j,\pm n}^{(p)}\right)\sqrt[]{\left(\Gamma_{j,\pm n}^{(p)}\right)^2+\left(\Sigma_{j,\pm n}^{(p)}\right)^2},\\
\bar{\phi}_{j,\pm n}^{(p)}&=\tan^{-1}\left(- \frac{\Sigma_{j,\pm n}^{(p)}}{\Gamma_{j,\pm n}^{(p)}}\right),
\end{align}
\end{subequations}
 such that 
%By employing the trigonometric identity for $\cos(a+b)=\cos(a)\cos(b)-\sin(a)\sin(b)$, we can further simplify Eq.(9) which then becomes,
\begin{widetext}
\begin{align} \label{ht_1PN}
h(t)=& \, \frac{G\,m\,\eta}{c^2\, D_L} \,x\, \bigg\{\bigg[ \sum_{j=1}^8 \alpha_{j,-2}^{(0)} \cos(j\phi-(j-2)\phi'+\bar{\phi}_{j,-2}^{(0)}) + \sum_{j=1}^6 \alpha_{j,0}^{(0)} \cos(j\phi-j\phi'+\bar{\phi}_{j,0}^{(0)}) \nonumber \\ & + \sum_{j=1}^4 \alpha_{j,+2}^{(0)} \cos(j\phi-(j+2)\phi'+\bar{\phi}_{j,+2}^{(0)})\bigg] + x^{0.5}\,\delta \bigg[ \sum_{j=1}^7  \alpha_{j,-1}^{(0.5)}\cos(j\phi-(j-1)\phi'+\bar{\phi}_{j,-1}^{(0.5)}) \nonumber \\ & + \sum_{j=1}^5 \alpha_{j,+1}^{(0.5)} \cos(j\phi-(j+1)\phi'+\bar{\phi}_{j,-1}^{(0.5)}) + \sum_{j=1}^9 \alpha_{j,-3}^{(0.5)} \cos(j\phi-(j-3)\phi'+\bar{\phi}_{j,-3}^{(0.5)})  \nonumber \\ & + \sum_{j=1}^3  \alpha_{j,+3}^{(0.5)} \cos(j\phi-(j+3)\phi'+\bar{\phi}_{j,+3}^{(0.5)}) \bigg] +\,x \bigg[\sum_{j=1}^8 \alpha_{j,-2}^{(1)} \cos(j\phi-(j-2)\phi'+\bar{\phi}_{j,-2}^{(1)}) \nonumber \\ & + \sum_{j=1}^4 \alpha_{j,+2}^{(1)} \cos(j\phi-(j+2)\phi'+\bar{\phi}_{j,+2}^{(1)}) + \sum_{j=1}^6 \alpha_{j,0}^{(1)} \cos(j\phi-j\phi'+\bar{\phi}_{j,0}^{(1)}) \nonumber \\ & + \sum_{j=1}^{10} \alpha_{j,-4}^{(1)} \cos(j\phi-(j-4)\phi'+\bar{\phi}_{j,-4}^{(1)}) + \sum_{j=1}^2 \alpha_{j,+4}^{(1)} \cos(j\phi-(j+4)\phi'+\bar{\phi}_{j,+4}^{(1)}) \bigg] \bigg\}.
\end{align}
\end{widetext}

A cursory look at the above equation may give the impression that 
the summation indices in various sums are terminated in an arbitrary manner.
Interestingly, we find a possible way to predict the maximum value that j index can take in each of the above summations.
This is related to the argument of $\phi'$ in each of these cosine series.
We infer that the argument of $\phi'$ can take a maximum value of six 
as we are restricting eccentricity contributions to sixth order in $e_t$.
This ensures that $j$ index can take maximum values of $8,6$ and $4$ 
at the Newtonian order in the above expression.
In other words, $j_{\rm max}$ in the above expression is given such that
$j_{\rm max} \pm n = 6$ where $\pm n$ value arises from the the argument of $\phi'$ variable in various summations.
It is easy to see that the above relation holds true even at 0.5 and 1PN orders
and it provides a natural check on the structure of these higher order PN contributions to $h(t)$.

To obtain GW response function for eccentric inspirals, we need to incorporate 
temporal evolution in $\omega, e_t, \phi$ and $\phi'$, given by our earlier 
listed 3PN-accurate differential equations.
The fact that we require to solve the above four coupled differential equations numerically ensures that
our approach to obtain ready-to-use $h(t)$ will be computationally expensive. This is clearly one of the motivation to obtain fully analytic $\tilde{h}(f)$ for compact binaries inspiraling along moderately eccentric orbits. Fortunately, we are in a position to compute analytic amplitude corrected $\tilde{h}(f)$ that incorporates 3PN-accurate Fourier phase while keeping eccentricity contributions accurate to sixth order in $e_0$ at every PN order. \\

 \section{\label{sec:level2} Analytic $\tilde{h}(f)$ for eccentric inspirals with 1PN amplitude corrections}

We first provide a detailed description of our approach to compute analytic Fourier transform of the restricted time domain inspiral family, given by Eq.~(\ref{Eq_h_t_0f}).
This will be followed by computing $\tilde h(f)$ associated 
with Eq.~(\ref{ht_1PN}). Preliminary data analysis implications of our analytic $\tilde{h}(f)$ are probed in Sec.~\ref{sec:level2A}.  
  
\subsection{Approach to compute Fourier transform of $h(t)$ for compact binaries inspiraling along precessing eccentric orbits } \label{sec:3A}

We begin by listing the expanded version of our quadrupolar order $h(t)$,
namely Eq.~(\ref{Eq_h_t_0f}) with $\mathcal{O}(e_t^4)$  
eccentricity contributions as
\begin{widetext}
\begin{align}
\label{Eq_3.1}
h(t)=& \, \frac{G\,m\,\eta}{c^2\,D_L} x \bigg\{\bigg[\alpha_{1,-2}^{(0)}\cos\left(\phi+\phi'+\bar{\phi}_{1,-2}^{(0)}\right)+ \alpha_{2,-2}^{(0)}\cos\left(2\phi+\bar{\phi}_{2,-2}^{(0)}\right)+\alpha_{3,-2}^{(0)}\cos\left(3\phi-\phi'+\bar{\phi}_{3,-2}^{(0)}\right) \nonumber \\ & + \alpha_{4,-2}^{(0)}\cos\left(4\phi-2\phi'+\bar{\phi}_{4,-2}^{(0)}\right) + \alpha_{5,-2}^{(0)}\cos\left(5\phi-3\phi'+\bar{\phi}_{5,-2}^{(0)}\right) + \alpha_{6,-2}^{(0)}\cos\left(6\phi-4\phi'+\bar{\phi}_{6,-2}^{(0)}\right)\bigg] \nonumber \\ & + \bigg[\alpha_{1,0}^{(0)} \cos\left(\phi-\phi'+\bar{\phi}_{1,0}^{(0)}\right) + \alpha_{2,0}^{(0)} \cos\left(2\phi-2\phi'+\bar{\phi}_{2,0}^{(0)}\right)+\alpha_{3,0}^{(0)} \cos\left(3\phi-3\phi'+\bar{\phi}_{3,0}^{(0)}\right)+\alpha_{4,0}^{(0)} \cos\left(4\phi-4\phi'+\bar{\phi}_{4,0}^{(0)}\right)\bigg]  \nonumber \\ & + \bigg[\alpha_{1,+2}^{(0)} \cos\left(\phi-3\phi'+\bar{\phi}_{1,+2}^{(0)}\right)+\alpha_{2,+2}^{(0)} \cos\left(2\phi-4\phi'+\bar{\phi}_{2,+2}^{(0)}\right)\bigg]\bigg\}.
\end{align}
\end{widetext}
%A close inspection reveals that above equation is an expanded form of Eq.~\ref{Eq_h_t_0f}. 
Clearly, we see three distinct square brackets that contain  three cosine functions with explicitly time dependent arguments, namely $j\phi-(j-2)\phi'$, $j\phi-j\phi'$ and $j\phi-(j+2)\phi'$. 
%Note that GW response function for inspiraling eccentric binaries is obtained by numerically imposing 
%$\omega(t)$ and $e_t(t)$ with the help of 
%fully 3PN-accurate  $ \dot \omega$ and $\dot e_t$, provided by Eqs.~(\ref{Tevolve_3}) and (\ref{Tevolve_4}).
Note that  $\alpha_{j,\pm n}^{(0)}$ and 
$ \bar{\phi}_{j,\pm n}^{(0)}$ experience implicit temporal evolution due to the GW emission induced variations to $\omega$ and $e_t$.
The main reason for displaying the above equation is to show explicitly 
how the periastron advance, defined by $\phi'$, influences the harmonic structure of $h(t)$ in comparison with Eq.~(4.21) of Ref.~\cite{YABW} or our Eq.~(\ref{Eq_hN}).
%dependence of {\color{red}periastron advance on the harmonic 
%structure of $h(t)$}. We see that -----

  We  obtain an analytic Fourier domain version of the above equation 
with the help of %\textit
the {Stationary Phase Approximation}, detailed in \cite{BO_book}.
How this approach can be employed to compute 
$\tilde{h}(f)$ for compact binaries spiraling along Keplerian
eccentric orbits can be found in  Sec.~IV of Ref.~\cite{YABW}.
 This approximation is quite appropriate for us as it provides a prescription to compute the asymptotic behavior of
generalized  cosine time series, as given by our Eq.~(\ref{Eq_3.1}).
Without loss of any generality, we may write such a time series as 
 \begin{align}
S(t)= s(t)\cos(l\phi(t))\,,
\end{align}
where $l>0$ and as expected 
 $S(t)$ should be a product of slowly varying amplitude $s(t)$ and a rapidly varying cosine function 
 with argument $l\phi(t)$.
Due to the virtue of Riemann-Lebesgue lemma, as noted in Ref.~\cite{BO_book}, the 
Fourier transform of $S(t)$ becomes 
\begin{align}
S_f(f)=\frac{1}{2}\int_{-\infty}^{\infty}s(t)e^{if(2\pi t-l\phi(t)/f)}dt\,.
\end{align}
It is not difficult to gather that the argument of the exponential function vanishes at the stationary point $t_0$ such that 
 $l\dot{\phi}(t_0)=2\pi f$. This allows us to invoke the approach of SPA to obtain the 
 asymptotic behaviour of $S_f(f)$ by the following expression:
\begin{align}
S_f(f)=&\, s(t_0)e^{-i\Psi(t_0)\pm i \pi/(2\times2)}\left[\frac{2!}{f|\Psi^{(2)}(t_0)|}\right]^{\frac{1}{2}}\frac{\Gamma \left(1/2 \right)}{2}  \nonumber \\
=& \,  \frac{s(t_0)}{2\,\sqrt[]{l\dot{F}(t_0)}}e^{-i(\Psi(t_0)\mp\pi/4)}\,,  \label{eq:SPA}
\end{align}
where the Fourier phase is  defined as $$\Psi(t)\coloneqq -2\pi f t\,+\,l\phi(t)$$
Note that  $F(t)=\dot{\phi}(t)/2\pi$
and therefore its value at the stationary point should be $F(t_0)= f/l$.
Interestingly, a rather identical computation can be done to obtain 
the Fourier transform of a similar sinusoidal time series to be
$i\,S_f(f)$. \\

 To make operational the above expression for $S_f(f)$, we require an explicit expression for the above defined Fourier phase at the stationary point $t_0$, namely
  \begin{align} \label{eq:Psi}
\Psi(t_0) \coloneqq -2\pi f t_0\,+\,l\phi(t_0)
  \end{align} 
This is done by defining $\tau=F/\dot{F}$ such that $\phi(F)$ and $t(F)$ become
%\begin{subequations}
\begin{align} 
\label{eq:phi} \phi(F)=&\, \phi_c + 2\pi \int^{F} \tau'dF'\,,\\
\label{eq:t} t(F)=&\, t_c+\int^{F}\frac{\tau '}{F'}dF'\,,
\end{align}
%\end{subequations}
where $\phi_c$ and $t_c$ are the orbital phase and time at coalescence. In the present context,  $\tau$ is defined using our 3PN-accurate expression for $\dot{\omega}$ given by Eq.~(\ref{Tevolve_3}).
Additionally, we require 3PN-accurate $e_t (\omega, \omega_0, e_0)$ 
expression, namely 3PN extension of Eq.~(\ref{16}), for  computing these integrals  analytically.
The expression for $\Psi[F(t_0)]$ obtained using Eq.~(\ref{eq:phi}) and (\ref{eq:t}) in (\ref{eq:Psi}) may be written as 
\begin{align} \label{Eq_3_2}
\Psi_l[F(t_0)]=\,l\phi_c-2\pi ft_c+2\pi\int^{F(t_0)} \tau' \left(l- \frac{f}{F'}\right)dF'\,,
\end{align}
where $F(t_0) = f/l$. In the present context, 
 we need to evaluate the above integral at a point of time where 
 the orbital frequency is related to the Fourier 
frequency by  $F(t_0) = f/l$.
A close inspection of Eq.~(\ref{Eq_3.1}) reveals that our expression for the quadrupolar order time domain response function 
 is structurally similar to the above displayed cosine time series and therefore we can easily adapt these 
 results to obtain the Fourier transform of our quadrupolar order
 $h(t)$.
 %(Eq.(\ref{Eq_3.1})).
 However, the SPA based $\tilde{h}(f)$ will have contributions from a number of distinct stationary points. 
This is primarily due to the fact that Eq.~(\ref{Eq_3.1}) consists of cosine functions of three different arguments, namely
$ j\, \phi - (j+2)\, \phi', j\, \phi - (j -2)\, \phi' $ and $ j\,\phi - j\,\phi'$. Note that there are only 
three distinct types of cosine arguments as we restricted our attention to the quadrupolar order GW response function for eccentric inspirals.
However, we infer from our 1PN-accurate $h(t)$, given by Eq.~(\ref{ht_1PN}), that there are {\it nine} distinct types of cosine functions 
with arguments $j\phi-(j\pm n)\phi'$ where $n=0,1,2,3,4$.
The associated {\it nine} stationary points 
$t^{\pm n}$ are computed by demanding that  $\dot \Psi^{\pm n}(t^{\pm n})=0$, where $\Psi^{\pm n}(t) \coloneqq -2\pi f t\,+\,j\phi-(j \pm n)\phi'$.
\\
 For computing Fourier transform of
 Eq.(\ref{Eq_3.1}), we solve $\dot{\Psi}^{\pm n}(t^{\pm n})= 0 $ to get the relevant stationary points  and this leads to 
\begin{align}
-2\pi f +j\dot{\phi}-(j \pm n)\dot{\phi'} &=0 \,,
\end{align}
%\dot{\Psi}^{\pm n}(t^{\pm n})=& \, 0 \Rightarrow 
%\dot{\phi}=&\, \omega\,=\,(1+k)\,\dot{l}\,=\,(1+k)\,n \\
%\dot{\phi'}=&\, k\,\dot{l}\,=\,k\,\frac{\dot{\phi}}{1+k}\,=\,\dot{\phi}(k-k^2+k^3)
where $\dot \phi = N(1+k)$ and this by definition is $\omega$.
%gives $ \dot \phi = \omega$.
The treatment of $\dot \phi'$ requires PN approximation 
as $ \dot \phi'$ equals  $k\,N$ ( this is because
$\phi' = k\, l$).
We need to express $k\,N$ 
in terms of $\omega$ and this leads to $\dot \phi'= \omega\, k/( 1+k)$ 
as $\omega = N( 1+ k) $.
For computing Fourier phase analytically, we 
 express $\dot \phi'$ as $\omega\, k^{(6)}_{(3)} $, where 
$k^{(6)}_{(3)} $ stands for 
the 3PN-accurate 
expression for $k/(1+k)$ that incorporates $e_t$ contributions accurate to ${\cal O}(e_t^6)$.
%{\color{red}{It is fairly straightforward to obtain such an expression from %Eq.~() in Ref.~\cite{KG06} and 
%the 3PN accurate relation connecting $n$ and $\omega$}}.
The resulting expression reads
\begin{widetext}
\begin{align} \label{k63}
k^{(6)}_{(3)} =&\, x\bigg\{3 \bigg[1+e_t^2+e_t^4+e_t^6\bigg]\bigg\}+x^2\bigg\{\frac{9}{2}-7 \eta +\bigg[\frac{87}{4}-\frac{41 \eta }{2}\bigg] e_t^2+\bigg[39-34 \eta\bigg] e_t^4+\bigg[\frac{225}{4}-\frac{95 \eta }{2}\bigg] e_t^6\bigg\}+x^3\bigg\{\frac{27}{2} \nonumber \\ &+\bigg(-\frac{481}{4}+\frac{123 \pi ^2}{32}\bigg) \eta +7 \eta ^2+\bigg[\frac{519}{4}+\bigg(-\frac{2037}{4}+\frac{1599 \pi ^2}{128}\bigg)\eta +61 \eta ^2\bigg] e_t^2+\bigg[\frac{2811}{8}+\bigg(-1174+\frac{3321 \pi ^2}{128}\bigg) \eta \nonumber \\ &+\frac{1361}{8}\eta ^2\bigg] e_t^4+\bigg[\frac{10779}{16}+\bigg(-\frac{16901}{8}+\frac{2829 \pi ^2}{64}\bigg) \eta +\frac{2675}{8} \eta ^2\bigg] e_t^6\bigg\}.
\end{align}
\end{widetext}
With the help of these inputs,  the stationary points 
$t^{\pm n}$, where $\dot \Psi^{\pm n}(t^{\pm n})$ vanish, are given by 
\begin{align*}
\left(j-(j \pm n)\, k^{(6)}_{(3)}\right)\,\dot{\phi}(t^{\pm n})\, &=2\,\pi\,f\,.
\end{align*}
In other words, the stationary phase condition is given by
\begin{align}\label{SPA_1}
F(t^{\pm n})\,=\, \frac{f}{ \left(j-(j \pm n)\, k^{(6)}_{(3)}\right)}\,.
\end{align}
 Rewriting $\Psi^{\pm n}(t) \coloneqq -2\pi f t+j\phi-(j \pm n)\phi'$ using relation between $\phi'$ and $\phi$ $\left(\phi'=  k^{(6)}_{(3)} \phi \right)$ gives $\Psi^{\pm n}(t) \coloneqq -2\pi f t+\left(j-(j \pm n)k^{(6)}_{(3)}\right)\phi$. We are now in a position to obtain analytic PN-accurate expressions for 
Fourier phases, associated with these stationary  points.
With Eq.~(\ref{eq:phi}) and (\ref{eq:t}), our Eq.~(\ref{Eq_3_2}) becomes
\begin{widetext}
\begin{align}
\label{Eq_3_5}
\Psi^{\pm n}_j[F(t^{\pm n})]=\, \left(j-(j\pm n)k^{(6)}_{(3)}\right) \phi_c - 2\pi f t_c + 2\pi\int^{F(t^{\pm n})} \tau' \left( j-(j \pm n)\,k^{(6)}_{(3)} - \frac{f}{F'}\right)dF'\,.
\end{align}
\end{widetext}
Note that $n$ takes values $0$ and $2$ as we are dealing with quadrupolar order GW response function %{\color{red}that incorporates $\mathcal{O}(e_t^4)$ eccentricity contributions},
given by
Eq.(\ref{Eq_3.1}).
However, $n$ varies from $0$ to $4$ if the underlying GW response function contains 1PN-accurate amplitude corrections that include
at each PN order eccentricity corrections  accurate to 
{${\cal O}(e_t^6)$}. Further, 
we do not display here  3PN-accurate expression for $\tau$ 
that includes the leading order $e_t$ corrections,
 listed as Eqs.~(6.7a) and (6.7b) in Ref.~\cite{MFAM16}.
 % and therefore we do not display it here.
However, we do list below the explicit 3PN-accurate $\Psi_j^{\pm n}[F(t^{\pm n})]$ that incorporates 
leading order $e_0$ contributions at each PN order: 
\begin{widetext}
\begin{align}   \label{psi_e02}
\Psi_j^n =&\, \left(j-(j+n)k^{(6)}_{(3)}\right) \phi_c - 2\pi f t_c - \frac{3 \, j}{256\, \eta \, x^{5/2}} \bigg\{1-\frac{2355}{1462} e_0^2 \chi ^{-19/9}+ x\bigg[-\frac{2585}{756}-\frac{25 n}{3 j}+\frac{55}{9}\eta \nonumber \\ & +\bigg(\bigg(\frac{69114725}{14968128}+\frac{1805 n}{172 j}-\frac{128365}{12432} \eta\bigg) \chi^{-19/9}+\bigg(-\frac{2223905}{491232}+\frac{15464 }{17544}\eta\bigg) \chi ^{-25/9}\bigg) e_0^2 \bigg] \bigg\}
+x^{3/2}\bigg[-16 \pi \nonumber \\ & +\bigg(\frac{65561 \pi}{4080}\chi ^{-19/9}-\frac{295945 \pi}{35088}\chi ^{-28/9}\bigg) e_0^2\bigg]+x^2\bigg[ -\frac{48825515}{508032}-\frac{31805 n}{252 j}+\bigg(\frac{22105}{504}-\frac{10 n}{j}\bigg) \eta +\frac{3085}{72}\eta^2 \nonumber \\ & +\bigg(\bigg(\frac{115250777195}{2045440512}+\frac{323580365 n}{5040288 j}+\bigg(-\frac{72324815665}{6562454976}+\frac{36539875 n}{1260072 j}\bigg)\eta -\frac{10688155}{294624}\eta^2\bigg) \chi^{-19/9} \nonumber \\ & +\bigg(\frac{195802015925}{15087873024}+\frac{5113565 n}{173376 j}+\bigg(-\frac{3656612095}{67356576}-\frac{355585 n}{6192 j}\bigg) \eta +\frac{25287905}{447552}\eta^2\bigg) \chi^{-25/9}+\bigg(\frac{936702035}{1485485568} \nonumber \\ & +\frac{3062285}{260064}\eta-\frac{14251675}{631584}\eta^2\bigg)\chi^{-31/9}\bigg) e_0^2\bigg]+x^{5/2}\bigg[\frac{14453 \pi }{756}-\frac{32 \pi n}{j}-\frac{65 \pi }{9} \eta -\bigg(\frac{1675}{756}+\frac{160 n}{3 j}+\frac{65}{9}\eta\bigg) \pi \log\left(\frac{f}{j}\right) \nonumber \\ & +\bigg(\bigg(-\frac{458370775 \pi }{6837264}-\frac{4909969 \pi  n}{46512 j}+\frac{15803101 \pi  \eta }{229824}\bigg) \chi ^{-19/9}+\bigg(\frac{185734313 \pi
   }{4112640}-\frac{12915517 \pi  \eta }{146880}\bigg) \chi ^{-25/9} \nonumber \\ & +\bigg(\frac{26056251325 \pi }{1077705216}+\frac{680485 \pi  n}{12384 j}-\frac{48393605 \pi  \eta }{895104}\bigg) \chi ^{-28/9}+\bigg(-\frac{7063901 \pi }{520128}+\frac{149064749 \pi  \eta }{2210544}\bigg) \chi^{-34/9}\bigg)e_0^2\bigg]\nonumber \\ &+x^{3}\bigg[\frac{13966988843531}{4694215680}+\frac{257982425 n}{508032 j}-\frac{640 \pi ^2}{3}-\frac{6848 \gamma
   }{21}+\bigg(-\frac{20562265315}{3048192}-\frac{2393105 n}{1512 j}+\frac{23575 \pi ^2}{96}\nonumber \\ &+\frac{1845 \pi ^2 n}{32 j}\bigg) \eta +\bigg(\frac{110255}{1728}+\frac{475 n}{24 j}\bigg) \eta ^2-\frac{127825 \eta ^3}{1296}-\frac{13696 \log (2)}{21}-\frac{3424 \log (x)}{21} \nonumber \\ & +\bigg(\bigg(\frac{4175723876720788380517}{5556561877278720000}+\frac{534109712725265 n}{2405438042112 j}-\frac{21508213 \pi ^2}{276480}-\frac{734341 \gamma}{16800} +\bigg(-\frac{37399145056383727}{28865256505344}\nonumber \\ &-\frac{1219797059185 n}{2045440512 j}+\frac{12111605 \pi ^2}{264192}+\frac{639805 n \pi^2}{22016 j}\bigg) \eta +\bigg(-\frac{159596464273381}{1718170030080} +\frac{43766986495 n}{1022720256 j}\bigg) \eta ^2-\frac{69237581}{746496}\eta^3 \nonumber \\ &-\frac{9663919 \log (2)}{50400}+\frac{4602177 \log (3)}{44800}-\frac{734341 \log (x)}{33600}\bigg)\chi ^{-19/9}+\bigg(\frac{326505451793435}{2061804036096}+\frac{916703174045 n}{5080610304 j}\nonumber \\ &-\bigg(\frac{13467050491570355}{39689727694848}+\frac{9519440485 n}{35282016 j}\bigg) \eta -\bigg(\frac{2186530635995}{52499639808} +\frac{7198355375 n}{45362592 j}\bigg) \eta ^2+\frac{2105566535 }{10606464}\eta^3\bigg) \chi ^{-25/9}\nonumber \\ &+\frac{24716497 \pi ^2 }{293760}\chi ^{-28/9}+\bigg(-\frac{82471214720975}{45625728024576}-\frac{2153818055 n}{524289024 j} +\bigg(-\frac{48415393035455}{1629490286592}-\frac{119702185 n}{1560384 j}\bigg) \eta \nonumber \\ &+\bigg(\frac{906325428545}{6466231296}+\frac{32769775 n}{222912 j}\bigg) \eta ^2-\frac{2330466575}{16111872} \eta ^3\bigg) \chi ^{-31/9} +\bigg(-\frac{4165508390854487}{16471063977984}-\frac{96423905 \pi ^2}{5052672}\nonumber \\ &\qquad+\frac{2603845 \gamma}{61404}+\bigg(-\frac{1437364085977}{53477480448}+\frac{3121945 \pi ^2}{561408}\bigg) \eta +\frac{4499991305 \eta ^2}{636636672}+\frac{2425890995 \eta^3}{68211072}+\frac{1898287 \log (2)}{184212}\nonumber \\ &\qquad +\frac{12246471 \log (3)}{163744}+\frac{2603845 \log (x)}{122808}-\frac{2603845 \log (\chi)}{184212}\bigg)\chi ^{-37/9}\bigg)e_0^2\bigg]\,.
\end{align}
\end{widetext}
%where $---$.
%\\

%Upto what PN \Psi we want to put? and what is the correct notation for it? 
 A few comments are in order. To obtain the circular limit, we require to impose $n=j$ in $j\phi-(j-n)\phi'$ and let $e_0=0$.
 This is indeed due to the fact that  $k^{(6)}_{(3)}$  does not go to zero in the circular limit. Additionally,
we have verified that the resulting $\Psi^{-2}_{2} (f) $ expression in the $e_0 \rightarrow 0$ limit is identical to 
3PN accurate version of Eq.~(6.26) in Ref.~\cite{MFAM16} while neglecting the spin contributions. 
It is natural to expect that the $\Psi^{0}_{j} (f) $ version of our above equation should be identical to Eq.~(6.26) of Ref.~\cite{MFAM16}.
This is because 
this equation indeed provided quadrupolar $\tilde{h}(f)$ with 3PN-accurate Fourier phase while incorporating leading order $e_0$ corrections at each PN order by extending the post-circular approach of Ref.~\cite{YABW}.
However, our expression for $\Psi^{0}_{j} (f) $ is not identical to Eq.~(6.26) of Ref.~\cite{MFAM16}.
This is because that effort did not incorporate the effect of periastron advance while obtaining analytic expression for their Fourier phase.
A close inspection of the $n=0$ version of our Eq.~(\ref{Eq_3_5}) reveals that it will still be influenced by our PN-accurate expression for $k^{(6)}_{(3)}$. This clearly shows that it is rather impossible to 
remove the effect of periastron advance from our Eq.~(\ref{Eq_3_5}).
Therefore, our Eq.~(\ref{psi_e02}) will be different from Eq.~(6.26) of Ref.~\cite{MFAM16} which, as noted earlier, neglected the effect of periastron advance.
The differences may be attributed to the 
physical fact that we are providing an analytic expression for $\tilde{h}(f)$ associated with compact binaries inspiraling along PN-accurate eccentric orbits. In contrast,  Ref.~\cite{MFAM16} models inspiral GWs from compact binaries spiraling in along Newtonian orbits though frequency evolution in both cases are fully 3PN accurate.
Additionally, we are unable to match with the 2PN order results of 
Ref.~\cite{THG} due to similar reasons.
We note in passing that the explicit 3PN-accurate ${\cal O}(e_0^4)$ contributions to $\Psi^{n}_{j}(f)$ and the associated
3PN accurate $e_t$ expression are provided in the Appendix~ \ref{appendixA} . 

We now employ fully the final result of SPA, namely  Eq.~(\ref{eq:SPA}),
to compute the Fourier transform of Eq.~(\ref{Eq_3.1}).
This gives us 
\begin{widetext}
\begin{align}
\tilde{h}[F(t_0)]=& \bigg(\frac{5\,\pi\,\eta}{384}\bigg)^{1/2} \frac{G^2m^2}{c^5D_L}\,\bigg(\frac{G\,m\,\pi\,2\,F(t_0)}{c^3}\bigg)^{-7/6}\,\frac{\left(1-e_t^2\right)^{7/4}}{\left(1+\frac{73}{24}e_t^2+\frac{37}{96}e_t^4\right)^{1/2}}\bigg\{\sum_{j=1}^6\alpha_{j,-2}^{(0)}\sqrt{\frac{2}{j}}e^{-i\bar{\phi}^{(0)}_{j,-2}[F(t_0)]}e^{-i(\Psi_j^{-2}+\pi/4)} \nonumber \\ & +\sum_{j=1}^4\alpha_{j,0}^{(0)}\sqrt{\frac{2}{j}}e^{-i\bar{\phi}^{(0)}_{j,0}[F(t_0)]}e^{-i(\Psi_j^{0}+\pi/4)} + \sum_{j=1}^2\alpha_{j,+2}^{(0)}\sqrt{\frac{2}{j}}e^{-i\bar{\phi}^{(0)}_{j,+2}[F(t_0)]}e^{-i(\Psi_j^{+2}+\pi/4)}\bigg\} \,, \label{hf_spa}   
\end{align}
\end{widetext}
where we have used the quadrupolar (Newtonian) order differential 
equation for the orbital frequency, available in Refs.~\cite{PM63,DGI},
to compute the amplitudes of $\tilde h[F(t_0)]$.
%evolution of eccentric binaries given by \cite{PM63}
%\begin{align}
%\frac{dF}{dt}=&\,\frac{48\,c^6\,\eta}{5\,\pi\,G^2\,m^2}\left(\frac{G\,m\,2\%,\pi\,F}{c^3}\right)^{11/3}\frac{\left(1+\frac{73}{24}e_t^2+\frac{37}{96}e_%t^4\right)}{\left(1-e_t^2\right)^{7/2}}.    
%\end{align}
Note that we require to employ the earlier defined 
stationary points 
to replace $F(t_0)$. In practice, we employ the 
unperturbed stationary points, namely  $F(t_0)=f/j$, while 
evaluating the amplitudes of $\tilde{h}(f)$.
 %has to be replaced using the SPA.{\color{red} We keep the amplitude of $\tilde{h}(f)$ at unperturbed stationary points such that $F(t_0)=f/j$}. 
 
  In what follows, we collect the above pieces together to display the quadrupolar order $\tilde{h}(f)$ that incorporates 
 fourth order orbital eccentricity contributions
 while including the effects due to
 3PN-accurate frequency, eccentricity evolution and periastron 
advance as
\begin{widetext}
\begin{align} \label{hf_newt}
\tilde{h}(f) &= \bigg(\frac{5 \pi  \eta }{384}\bigg)^{1/2}\frac{G^2 m^2}{c^5 D_L}\bigg(\frac{G m \pi  f}{c^3}\bigg)^{-7/6}\bigg\{\sum _{j=1}^6 \xi _{j,-2}^{(0)}\bigg(\frac{j}{2}\bigg)^{2/3} e^{-i \left(\Psi _j^{-2}+\frac{\pi }{4}\right)}+\sum _{j=1}^4 \xi _{j,0}^{(0)} \bigg(\frac{j}{2}\bigg)^{2/3} e^{-i \left(\Psi _j^0 + \frac{\pi }{4}\right)} \nonumber \\ & +\sum _{j=1}^2 \xi _{j,+2}^{(0)}\bigg(\frac{j}{2}\bigg)^{2/3} e^{-i \left(\Psi _j^{+2}+\frac{\pi }{4}\right)}\bigg\} , 
\end{align}
\end{widetext}
where the Fourier amplitudes $\xi_{j,\pm n}^{(0)}$ are now given by 
\begin{align}   \label{xi_newt}
\xi_{j,\pm n}^{(0)} &= \frac{\left( 1-e_t^2 \right)^{7/4}}{\left(1+\frac{73}{24}e_t^2+\frac{37}{96}e_t^4\right)^{1/2}} \alpha_{j,\pm n}^{(0)}\,e^{-\textit{i} \bar{\phi}_{j,\pm n}^{(0)}(f/j)}\,,
\end{align}
and  $n$ takes values 0 and 2.
A crucial expression that will be required to operationalize the above $\tilde{h}(f)$, namely 3PN-accurate expression for $e_t$ in terms of $e_0,x$ and $\chi$,  
is listed as Eq.~(\ref{appendixet}) in the Appendix. 
%\begin{align}
%e_t &=
%\end{align}
Note that the approach to obtain such an expression for $e_t$ is detailed in Ref.~\cite{THG} and briefly summarized in Sec.~\ref{sec:PostCirc_1}.
Finally, the fully 3PN-accurate expression for $\Psi^{n}_{j} (f) $ 
that incorporates fourth order orbital eccentricity contributions
at each PN order 
is displayed as Eq.~(\ref{appendixpsi})
in the Appendix. 
It should be noted that the approach of SPA
{ demands the evaluation of Fourier amplitudes, $\xi_{j,\pm n}$ and Fourier phases, $\Psi_j^{\pm n}$ at $F(t^{\pm n})\,=\,f/ \left(j-(j \pm n)\, k^{(6)}_{(3)}\right)$}.

 We have extended these calculations by including 1PN-accurate amplitude corrections to $h_{\times}$ and $h_+$
 with the help of Eqs.~(\ref{hpc}),(\ref{22}),(\ref{23}) and (\ref{24}).
 Additionally, we have included 
 initial eccentricity corrections, accurate to ${\cal O}(e_0^6)$, in our 3PN-accurate $e_t$ and $\Psi^{n}_{j} (f)$ expressions.
 We  note in passing that these expressions are available in the 
accompanying \texttt{Mathematica} file.
 The resulting expression for $\tilde{h}(f)$ may be symbolically written as
 \begin{widetext}
\begin{align}   \label{hf_1PN}
\tilde{h}(f) =&\, {\left(\frac{5 \pi  \eta }{384}\right)}^{1/2}\frac{G^2 m^2}{c^5 D_L}\left(\frac{G m \pi  f}{c^3}\right)^{-7/6}\bigg\{\bigg[\sum _{j=1}^6 \xi _{j,0}^{(0)} \left(\frac{j}{2}\right)^{2/3} e^{-i \left(\Psi _j^0+\frac{\pi }{4} \right)}+\sum _{j=1}^4 \xi _{j,+2}^{(0)} \left(\frac{j}{2}\right)^{2/3} e^{-i \left(\Psi _j^{+2}+\frac{\pi }{4}\right)} \nonumber \\ &  +\sum _{j=1}^8 \xi _{j,-2}^{(0)} \left(\frac{j}{2}\right)^{2/3} e^{-i \left(\Psi _j^{-2}+\frac{\pi }{4}\right)}\bigg]+\left(\frac{G m \pi  f}{c^3}\right)^{1/3} \delta \bigg[\sum _{j=1}^5 \xi _{j,+1}^{(0.5)}\left(\frac{j}{2}\right)^{1/3} e^{-i \left(\Psi _j^{+1}+\frac{\pi }{4}\right)} \nonumber \\ & +\sum _{j=1}^7 \xi _{j,-1}^{(0.5)}\left(\frac{j}{2}\right)^{1/3} e^{-i \left(\Psi _j^{-1}+\frac{\pi }{4}\right)}+\sum _{j=1}^3 \xi _{j,+3}^{(0.5)}\left(\frac{j}{2}\right)^{1/3} e^{-i \left(\Psi _j^{+3}+\frac{\pi }{4}\right)}+\sum _{j=1}^9 \xi _{j,-3}^{(0.5)}\left(\frac{j}{2}\right)^{1/3} e^{-i \left(\Psi _j^{-3}+\frac{\pi }{4}\right)}\bigg] \nonumber \\ &  +\left(\frac{G m \pi  f}{c^3}\right)^{2/3}\bigg[\sum _{j=1}^6 \xi _{j,0}^{(1)} e^{-i \left(\Psi _j^0 +\frac{\pi }{4}\right)}+ \sum _{j=1}^4 \xi _{j,+2}^{(1)} e^{-i \left(\Psi _j^{+2}+\frac{\pi }{4}\right)}+ \sum _{j=1}^8 \xi _{j,-2}^{(1)} e^{-i \left(\Psi _j^{-2}+\frac{\pi }{4}\right)} \nonumber \\ & + \sum _{j=1}^2 \xi _{j,+4}^{(1)} e^{-i \left(\Psi _j^{+4}+\frac{\pi }{4}\right)}+ \sum _{j=1}^{10} \xi _{j,-4}^{(1)} e^{-i \left(\Psi _j^{-4}+\frac{\pi }{4}\right)}\bigg]\bigg\} \,.
\end{align}
\end{widetext}
In the above expression, the Fourier amplitudes are given by
\begin{widetext}
\begin{align} \label{xi_1PN}
\xi_{j,\pm n}^{(p)} &= \bigg\{\frac{\left( 1-e_t^2 \right)^{7/4}}{\left(1+\frac{73}{24}e_t^2+\frac{37}{96}e_t^4\right)^{1/2}}+\frac{\left( 1-e_t^2 \right)^{3/4}}{10752\left(1+\frac{73}{24}e_t^2+\frac{37}{96}e_t^4\right)^{3/2}}\bigg[11888 + 14784 \eta - e_t^2 (87720 - 159600 \eta) \nonumber \\ & \qquad - e_t^4 (171038 - 141708 \eta) - e_t^6 (11717 - 8288 \eta)\bigg]\bigg\} \alpha_{j,\pm n}^{(p)}\,e^{-\textit{i} \bar{\phi}_{j,\pm n}^{(p)}},
\end{align}
\end{widetext}
where the superscript $p$ takes values $0,0.5$ and $1$ in our amplitude corrected $\tilde{h}(f)$.
Further, we have used the  1PN-accurate differential equation for the orbital frequency while obtaining the Fourier amplitude expressions.
This expression, adaptable from Eqs.~(B8a) and (B9a) of Ref.~\cite{THG}, reads 
%For obtaining above $\tilde{h}(f)$ we have usedevolution of eccentric binaries given as,
\begin{widetext}
\begin{align}
\frac{dF}{dt}=&\,\frac{48\,c^6\,\eta}{5\,\pi\,G^2\,m^2}\left(\frac{G\,m\,2\,\pi\,F}{c^3}\right)^{11/3}\frac{\left(1+\frac{73}{24}e_t^2+\frac{37}{96}e_t^4\right)}{\left(1-e_t^2\right)^{7/2}} - \frac{743\,c^6\,\eta}{35\,\pi\,G^2\,m^2}\left(\frac{G\,m\,2\,\pi\,F}{c^3}\right)^{13/3}\frac{1}{\left(1-e_t^2\right)^{9/2}} \nonumber \\ &\bigg\{1  +\frac{924}{743}\eta+e_t^2\bigg(-\frac{10965}{1486}+\frac{9975}{743}\eta\bigg) +e_t^4\bigg(-\frac{85519}{5944}+\frac{35427}{2972}\eta\bigg)+e_t^6\bigg(-\frac{11717}{11888}+\frac{518}{743}\eta\bigg)\bigg\}\,.
\end{align}
\end{widetext}
The  explicit expressions for $e_t$ and $\Psi^{n}_{j}(f)$ that incorporate the next to leading order $e_0$ corrections at each PN order,
as noted earlier, are listed in the Appendix \ref{appendixA}.

%while the attached {\it Mathematica} file provides all these quantities accurate to ${\cal O}(e_0^6)$.
  We move on to contrast our approach with other attempts in the literature.
The Sec.~VI of Ref.~\cite{YABW}	indeed sketched a road map to include PN corrections to their Newtonian waveform family.
This road map included a suggestion to incorporate the effect of periastron advance into their quadrupolar order GW polarization states,
influenced by Ref.~\cite{DGI}. Their suggestion involves splitting the orbital phase evolution into two parts where one part remains
linear in the mean anomaly $l$ while the other part is periodic in $l$. These considerations influenced them to re-write our Eq.~(\ref{Eq_hN}) essentially to be 
\begin{align}
h(t)&=\, -\frac{G\,m\,\eta}{c^2\,D_L} x \sum_{j=1}^{10} \alpha_j \cos\bigg\{j\,l\,\left(1+k^{(6)}_{(1)}\right)+\phi_j \bigg\} \,,
\end{align}
where $k^{(6)}_{(1)}$ stands for the 1PN accurate expression for $k$, given by $3\,x/(1-e_t^2)$,  expanded to the sixth order in $e_t$ (see our Eq.(\ref{k63})).
It is not difficult to see that the associated SPA based 
Fourier phase takes the following form:
%The above modification of our Eq.~(\ref{ht_1PN}) leads to the following SPA based Fourier phase 
\begin{align}
\Psi_j (F) &=\,\lambda\left[t\left(f/j\right)\right]-2\,\pi\,f\,t\left(f/j\right)
\,,
\end{align}
where
\begin{align}
\lambda\left[t\left(f/j\right)\right] &= \,j\,\phi_c+j\,\int^{f/j}\frac{\dot{\lambda}'}{\dot{F}'}dF' \,\\
t\left(f/j\right) &= t_c+\int^{f/j}\frac{dF'}{\dot{F}'}\,.
\end{align}
It turned out that  $\dot{\lambda}' \equiv \omega $ by construction.
%as the orbital angular frequency is related to the mean motion by $\omega = N(1 +k)$.
The use of $\omega$ in the above Fourier phase expression essentially  ensures that the suggestion of Ref.~\cite{YABW} leads
to what is detailed in Ref.~\cite{THG}. 
Note that Ref.~\cite{THG} provided $\tilde{h}(f)$ in terms of infinite set of harmonics with quadrupolar order amplitudes 
and 2PN-accurate Fourier phase. 
We observe that Ref.~\cite{YABW} indeed commented on the absence of side bands in their prescription in comparison 
with what was reported in Refs.~\cite{Moreno94,Moreno95} and suggested future investigations to clarify the issue.
In contrast, the present investigation 
employs Eqs.~(\ref{ht_1PN}) that explicitly incorporates the effect of periastron advance 
both in the amplitude and phase of GW polarization states, as detailed in Ref.~\cite{YS}.
The use of such an expression ensures that our analytic Fourier domain expression does indeed contain periastron advance 
induced frequency 
side bands. 
%Additionally, the periastron advance explicitly affects the computation of the Fourier phase under the SPA as 
%evident from our Sec.~\ref{sec:3A}. 
{ Additionally, Refs.~\cite{GKRF18,MKFV} employed 
the dominant order periastron advance induced decomposition 
of Fourier phases, associated with quadrupolar order 
gravitational waveform,  
 while exploring LISA and aLIGO 
relevant parameter estimation studies.
%of frequency domain gravitational waveform for precessing eccentric binaries with {\it quadrupolar} amplitude was carefully taken into account for parameter estimation studies carried out in Refs.~\cite{GKRF18,MKFV}. 
A close comparison of Eqs.~(B10) and (B11) of Ref.~\cite{MKFV} and Eqs.~(35) and (36) of Ref.~\cite{GKRF18} with our Eq.~(3.10) reveals
fairly identical expressions for the Fourier phases.
%a correct incorporation of precession effect in the Fourier phase of our analytic waveforms.
}
These considerations allowed us to state that our expression for $\tilde{h}(f)$, given by Eqs.~(\ref{hf_1PN}),(\ref{xi_1PN}),(\ref{appendixet}),(\ref{appendixpsi}), provides analytic 
PN-accurate Fourier domain templates 
for compact binaries inspiraling along PN-accurate precessing eccentric orbits.
We are now in a position to explore basic GW data analysis implications of our inspiral templates.

\subsection{ Preliminary GW data analysis implications
%Match computations to probe the relevance of higher order PN and $e_0$ corrections
}
 \label{sec:level2A}

 We employ the familiar  match  computations to probe basic GW data analysis implications of our PN-accurate inspiral templates.
Following Ref.~\cite{DIS98}, the match ${\cal M}(h_s,h_t)$ between 
members of two waveform classes, namely signal $h_s$ and template $h_t$, 
is computed by maximizing a certain overlap integral
$\mathcal{O}(h_s, h_t) $ with respect to the kinematic variables of the template waveform.
 In other words,
 \begin{align}
\label{Eq:match}
{\cal M}(h_s, h_t) &= \max_{t_0, \phi_0}\, \mathcal{O}(h_s, h_t)\,,
\end{align}
where $t_0$ and $\phi_0$ are the detector  
arrival time and the associated arrival phase of our template.
The overlap integral involves the interferometer-specific normalized inner product between members of 
$h_s$ and $h_t$ families; it reads 
\begin{align}
\label{Eq:innerproduct}
\langle  h_s |  h_t \rangle &= 4\, {\rm Re}\,  \int_{f_{\rm low}}^{f_{\rm high}} \, 
\frac{\tilde h_s^*(f)\, \tilde h_t(f)}{S_{\rm h}(f)} df \,,
\end{align}
where $\tilde h_s(f)$ and $\tilde h_t(f)$ are the Fourier transforms of the $h_s(t)$ and
$h_t(t)$ inspiral waveforms. Further,  $S_{\rm h}(f)$ denotes
the one-sided power spectral density of the detector noise. In the following,  we employ the
zero-detuned, high power (ZDHP) noise configuration of Advanced LIGO at design sensitivity \cite{LIGO_2010}.
% for the present ${\cal M}(h_e, h_a)$ computations.
In our ${\cal M}$ estimates,  we let $ f_{\rm low}$ be $20\,$Hz, corresponding to the lower cut-off frequency of Advanced LIGO.
The upper frequency limit $ f_{\rm high}$ is chosen to be the usual $f_{\rm LSO}=c^3/(G\, m\, \pi\, 6^{3/2})$ of the last stable circular orbit.
We have verified that orbital eccentricities of compact binaries reduce to well below $10^{-2}$ at ${f_{\rm high}} = f_{\rm LSO}$, thereby justifying the use of the last stable circular orbit frequency for the upper frequency limit.

We require additional steps to operationalize our inspiral templates
%, given by Eqs.~(\ref{hf_1PN}),(\ref{xi_1PN}),(\ref{appendixet}),(\ref{appendixpsi}), 
while performing the ${\cal M}$ computations.  Clearly, these waveform families should only be implemented 
within the physically allowed frequency intervals. This is to ensure that the many higher harmonics present in these 
waveform families do not cross the above listed upper frequency limit.
Influenced by Ref.~\citep{YABW},
we invoke the \textit{Unit Step  function} ($\Theta$)
 to operationalize our inspiral templates.
 This step function allows us to appropriately terminate the waveform
as $\Theta(y)=1$ for $y \geq 0$ and {\it zero} otherwise.
The structure of our quadrupolar amplitude inspiral family, given by Eqs.~(\ref{hf_newt}), 
compels us to invoke $\Theta$ functions such that 
\begin{widetext}
\begin{align} \label{eq:theta_hf}
\tilde{h}(f) = & \,\ {\bigg(\frac{5 \pi  \eta }{384}\bigg)}^{1/2}\frac{G^2 m^2}{c^5 D_L}\bigg(\frac{G m \pi  f}{c^3}\bigg)^{-7/6} \bigg \{ \sum _{j=1}^4 \xi _{j,0}^{(0)} \bigg(\frac{j}{2}\bigg)^{2/3} e^{-i \left(\frac{\pi }{4}+\Psi _j^0 \right)} \times \Theta\bigg[\bigg(j-j\,k^{(6)}_{(3)}\bigg)f_{LSO}-2f\bigg] \nonumber \\ & +\sum _{j=1}^2 \xi _{j,+2}^{(0)} \bigg(\frac{j}{2}\bigg)^{2/3} e^{-i \left(\frac{\pi }{4}+\Psi _j^{+2}\right)} \times \Theta\bigg[\bigg(j-(j+2)\,k^{(6)}_{(3)}\bigg)f_{LSO}-2f\bigg]+\sum _{j=1}^6 \xi _{j,-2}^{(0)} \bigg(\frac{j}{2}\bigg)^{2/3} e^{-i \left(\frac{\pi }{4}+\Psi _j^{-2}\right)} \nonumber \\ & \times \Theta\bigg[\bigg(j-(j-2)\,k^{(6)}_{(3)}\bigg)f_{LSO}-2f\bigg] \bigg\}\,.
\end{align}
\end{widetext} 
Note that we have appropriately shifted the upper frequency limits to ensure that higher  {\it harmonics} are suitably terminated. 
While implementing our $\tilde{h}(f)$ we have encountered the violation of the {stationary phase condition}, namely Eq.~(\ref{SPA_1}), at a few Fourier frequencies corresponding to lower harmonic indices ($j \sim 1,2$).
We infer that the periastron advance induced shift of these harmonics 
can lead to negative GW frequencies. Therefore, we have discarded such
Fourier components. Interestingly, Ref.~\cite{MG07} showed that these
harmonics provide negligible contributions to the GW power spectrum,
which may be used to justify our neglect of such Fourier components in the implementation of our waveform families.
The above steps ensure smoothly varying templates which we will use in the following to pursue match computations.
We provide three frequency series of the same length (corresponding to $h_s$ and $h_t$ inspiral families and the ZDHP noise power spectral density) and employ a routine from the free and open software package 
{\ttfamily PyCBC} \cite{alex_nitz_2019_2556644} to compute various ${\cal M}$ estimates.

\iffalse
Finally, we employ the {\it match} routine of 
a free and open software package 
{\ttfamily PyCBC} \cite{pycbc} to implement the above ${\cal M}$ estimates.
%https://pycbc.org/
%For pursuing match computations, we have utilised Pycbc's 'match' routine. A submodule of pycbc filter package, 'pycbc.filter.match' returns the match between two waveforms which is equivalent to the overlap maximized over time and phase. 
For obtaining various match estimates, 
we provide two frequency series corresponding to $h_s$ and $h_t$ inspiral families and 
 a third frequency series for the noise power spectral density to the  routine in frequency steps of $1/8$ Hz.
A submodule of {\ttfamily PyCBC}
filter package, namely {\it pycbc.filter.match}, returns a value  
between $0$ and $1$ and it specifies the amount of maximized overlap between the two inspiral waveforms.
For the ${\cal M}$ estimates of this Section, we vary $e_0$ values from 
$0$ to $0.4$ in steps of $0.0125$, giving $33$ sample points in the ${\cal M}-e_0$ plane.
We ensure that all input series have the same length as actual match implementation involves 
obtaining the Inverse Fast Fourier Transforms.
\fi

\begin{figure*}[htp]
\begin{center}
\includegraphics[width=0.5\textwidth, angle=0]{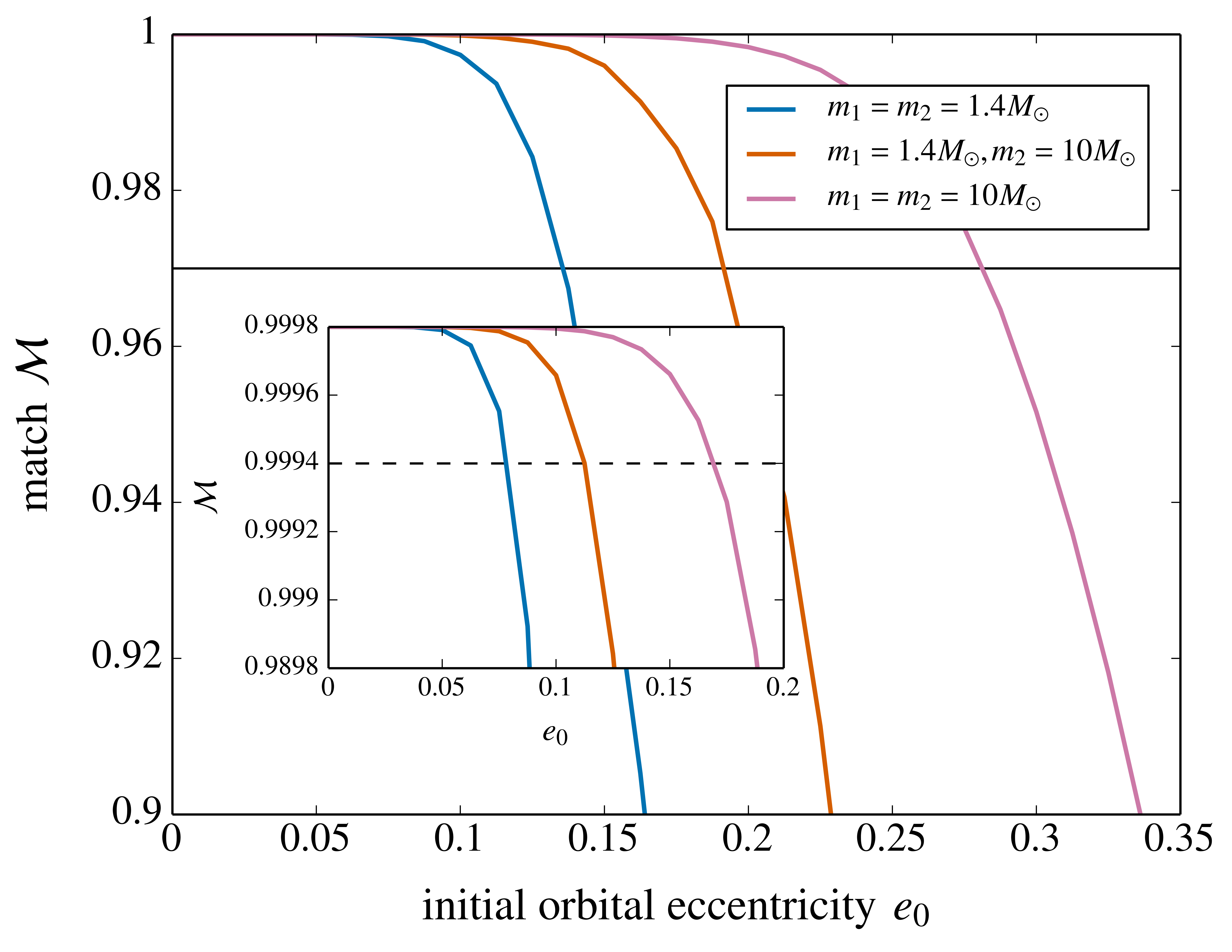}\\[0.1cm]
\end{center}
\caption{\label{fig:e02_e06} 
Matches between eccentric waveform models with different orders of eccentricity corrections. We are comparing waveforms that take only leading-order $\mathcal{O}(e_0^2)$ eccentricity corrections into account to those that include eccentricity corrections up to next-to-next-to leading order $\mathcal{O}(e_0^6)$. We consider three configurations of a NS and a BH with masses of $1.4 M_\odot$ and $10 M_\odot$, respectively: i.e., NS-NS (blue curve), NS-BH (orange curve) and BH-BH (pink curve) systems. The initial orbital eccentricity $e_0$ refers to the eccentricity of the binary system at 20 Hz. Given the same $e_0$, the effect of higher-order eccentricity corrections on the agreement between signal and template is strongly dependent on the total mass of the compact binary source. The solid black line denotes the threshold $\mathcal{M}=0.97$, associated with the effectualness of a model for GW detection and its faithfulness for source parameter estimation. The inset plot zooms into the region of parameter space where we can expect the effect of higher-order eccentricity corrections to become distinguishable from noise for SNR$\,= 30$, leading to systematic errors in parameter estimation; the dashed black line represents the indistinguishability criterion.}

%Plots showing match $\mathcal{M} (h_s,h_t)$ values between waveforms with only leading order eccentricity correction $(\mathcal{O}(e_0^2)\,terms)$ and with next-to-next to leading order corrections $(\mathcal{O}(e_0^6)\,terms)$ for different $e_0$ values. These plot show dependence of $\mathcal{M}$ on total mass of the system. The observed behavior is attributed to the time spent by the binary in aLIGO's sensitive frequency band which decreases with increasing total binary mass. In above plots, $10M_\odot$ and $1.4M_\odot$ represent a BH and a NS respectively in compact binary system with components $m_1$ and $m_2$. The solid black line represents $\mathcal{M}=0.97$ value associated with the faithfulness of $h_t$ with $h_s$.
\end{figure*}

\begin{figure*}[htp]
\begin{center}
\includegraphics[width=0.5\textwidth, angle=0]{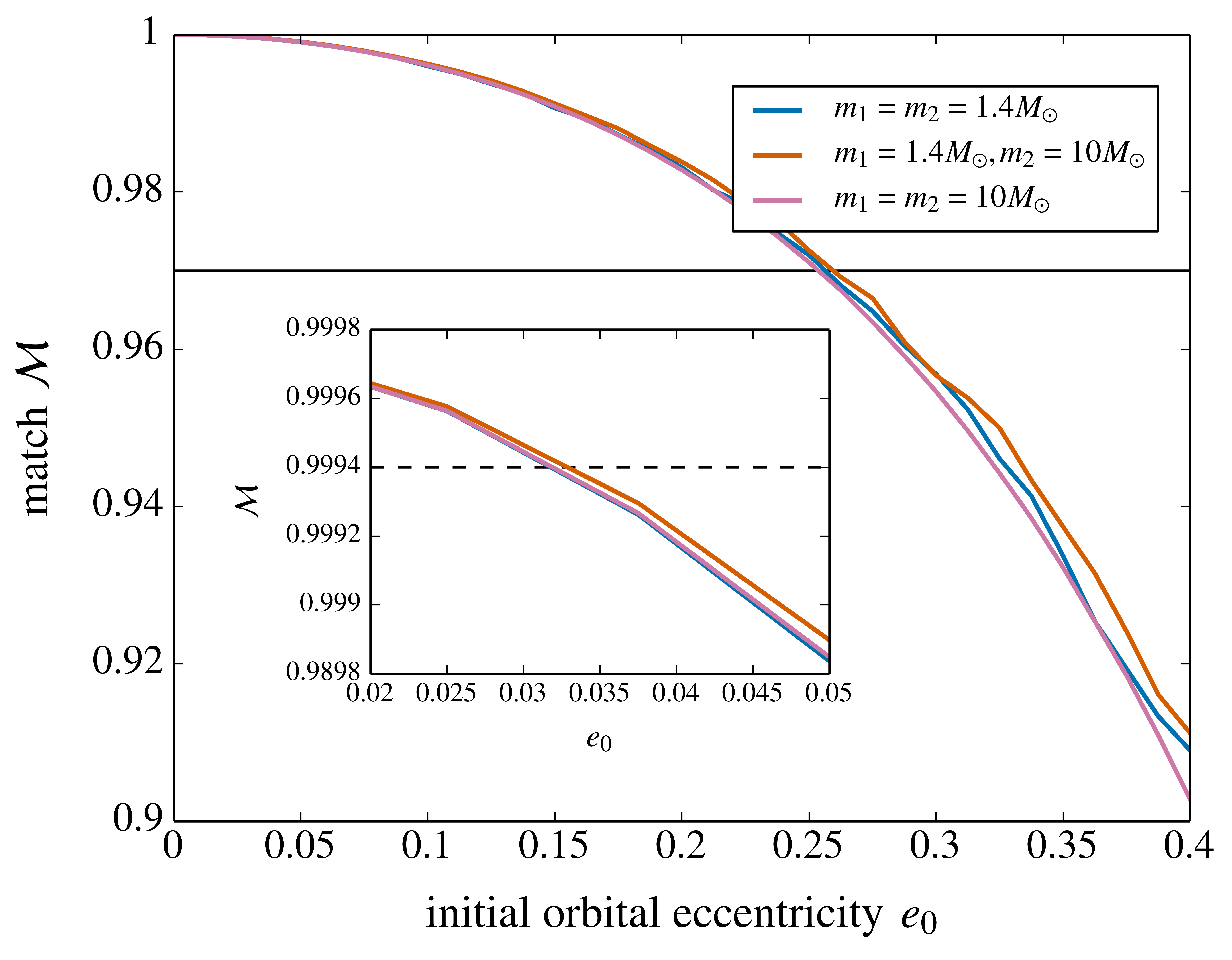}\\[0.1cm]
\end{center}
\caption{\label{fig:aop_waop} 
Matches between eccentric waveform models that include or neglect effects of periastron advance. We consider the same three configurations of binaries with NS and BH components as in Fig.~\ref{fig:e02_e06}: i.e., NS-NS (blue curve), NS-BH (orange curve) and BH-BH (pink curve) systems. The initial orbital eccentricity $e_0$ is again defined at the lower cut-off frequency $20\,$Hz. We infer that the significance of periastron advance effects for GW data analysis is rather independent of the total mass of the source. We interpret our results by considering the threshold $\mathcal{M}=0.97$ (represented by the solid black line) below which a waveform model should be considered ineffectual for detection and unfaithful for parameter estimation. In the inset plots, we highlight the parameter space of small eccentricities to probe the importance of systematic errors in parameter estimation due to waveform uncertainties. The dashed black line represents the distinguishable limit for a fiducial GW signal with SNR$\,=30$.
}
\end{figure*}

\begin{figure*}[htp]
\begin{center}
\includegraphics[width=0.5\textwidth, angle=0]{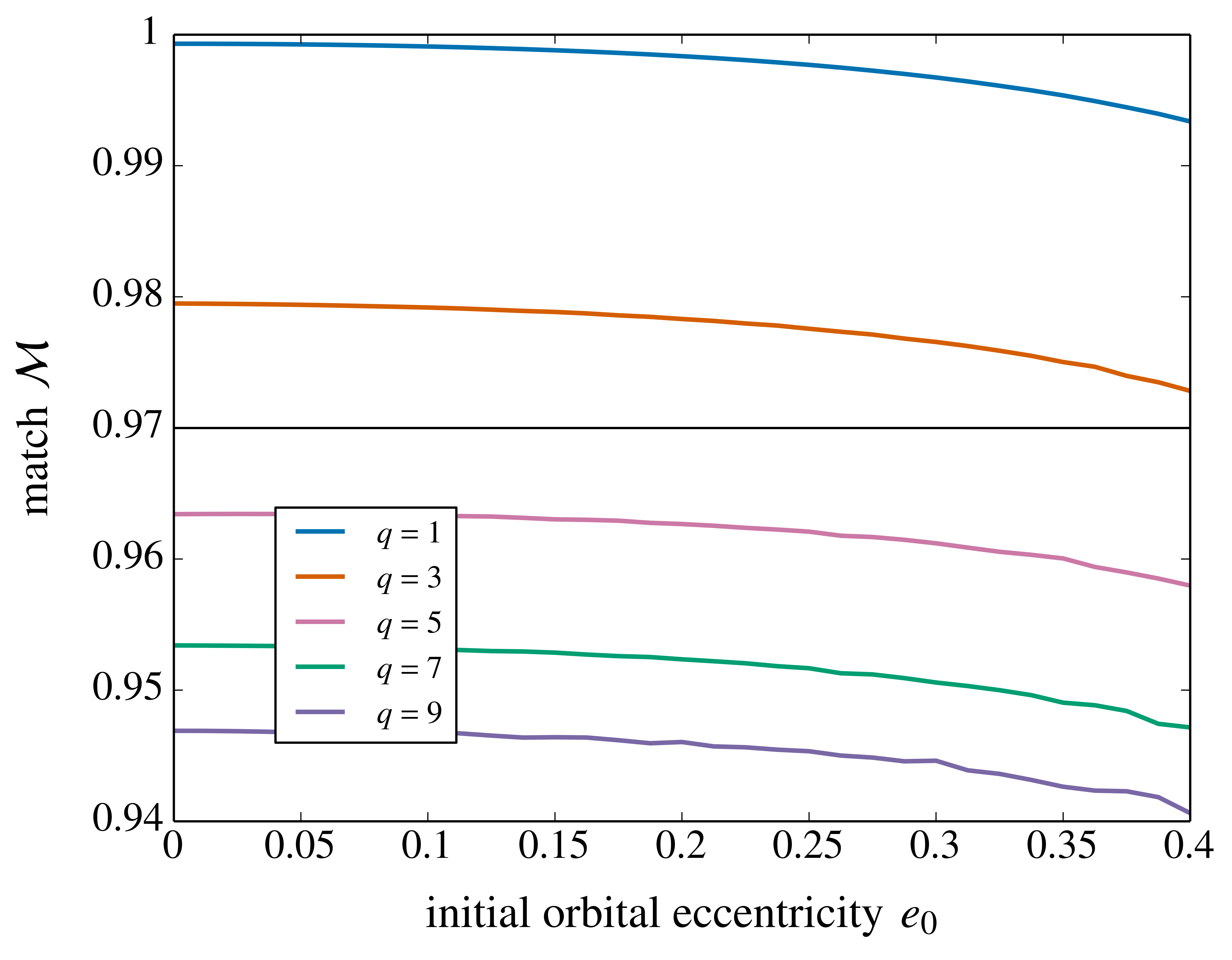}\\[0.1cm]
\end{center}
\caption{\label{fig:Newt_1PN}
Matches between eccentric waveform models with Newtonian and 1PN-accurate amplitudes. We consider compact binary systems with a total mass of $m=m_1+m_2$, with different choices for the mass ratio $q=m_1/m_2$. As expected, the effect of amplitude corrections on waveform faithfulness is largely independent of the orbital eccentricity $e_0$ at 20 Hz. Waveforms with Newtonian amplitudes are faithful representations of amplitude-corrected waveforms only if $q \leq 3$ (blue and orange curves); for higher mass ratios $q \geq 5$ (pink, green and purple curves) the match between waveforms with Newtonian and 1PN-accurate amplitudes falls below the threshold of $\mathcal{M}=0.97$ (denoted by the black line) even in the circular limit.
%Plots that probe variations of match $\mathcal{M}$ estimates between waveforms with Newtonian and 1PN accurate amplitudes as a function of $e_0$.
%These matches are performed by considering systems with different mass ratios $q=m_1/m_2$ while keeping the $m$ to be $20M_\odot$. The effect of amplitude corrections is seen to be independent of orbital eccentricity at  $f_0=20$Hz. Further, compact binaries with increasing mass ratio have poorer $\mathcal{M}$ values and amplitude corrected waveforms are relevant to faithfully model GWs from systems having unequal mass components.  
}
\end{figure*}

We qualify the implications of our match estimates on GW data analysis by considering the threshold $\mathcal{M}(h_s,h_t) \geq 0.97$, denoted in the presentation of results in Figs.~\ref{fig:e02_e06},~\ref{fig:aop_waop} and~\ref{fig:Newt_1PN} by solid black lines. This limit corresponds to a loss of less than $10\%$ of all signals in the matched filter searches.
In regions of parameter space where the computed matches are high, i.e., $\mathcal{M} \geq 0.97$, waveform models are generally considered both {\it effectual} templates for the detection of fiducial GW signals and reasonably {\it faithful} in the estimation of GW source parameters \cite{DIS98}.
However, even if $\mathcal{M}$ larger than $0.97$, certain errors in the model waveform (due to unmodeled effects of, e.g., eccentricity) may become {\it distinguishable} from noise at high signal-to-noise ratio (SNR) and can affect the accuracy of source parameter estimation. Negligible systematic errors in parameter estimation -- despite differences between the true signal waveform and the template model -- can be guaranteed only if $(h_s - h_t, h_s - h_t) < 1$, the so-called {\it indistinguishability criterion} \cite{creightonanderson}. In other words, such systematic errors in the estimated source parameters may become significant when the mismatch $1 - \mathcal{M}_c \geq 1/{\rm SNR}^2$ and clearly depend on the amplitude of the signal. In the following analysis, we let the signal-to-noise ratio of our fiducial GW signals be SNR$\,= 30$ (corresponding to the SNR of the binary neutron star inspiral GW170817) and probe the distinguishability of certain effects in our model waveforms for inspiraling eccentric binaries. In the inset plots of Figs.~\ref{fig:e02_e06} and~\ref{fig:aop_waop}, we zoom into those regions of parameter space where we can expect waveform uncertainties to become indistinguishable from noise for SNR$\,= 30$; the corresponding distinguishable limit $\mathcal{M}_c$ is represented by the dashed black lines.
 
 We first probe the importance of higher-order eccentricity
 corrections in the GW phasing. For this purpose, we let the signal family $h_s$ to be our quadrupolar-order $\tilde{h}(f)$, with a 3PN-accurate Fourier phase that 
 includes next-to-next-to leading order, ${\cal O}(e_0^6)$ eccentricity corrections at each PN order. The template family is given by a quadrupolar-order $\tilde{h}(f)$ in the low-eccentricity limit,
 incorporating only the leading-order, ${\cal O}(e_0^2)$ eccentricity contributions
 %${\cal O}(e_0^2)$ 
 in the 3PN-accurate Fourier phase. 
 We consider the traditional non-spinning compact binary sources relevant for Advanced LIGO: namely, binary neutron stars (NS-NS), NS-BH systems and binary black holes (BH-BH), with NS and BH components of $1.4\,M_{\odot}$ and $10\,M_{\odot}$, respectively.
 %The plots in 
  %Fig.~\ref{fig:e02_e06} are for the traditional aLIGO relevant non-spinning compact binaries, namely $1.4\,M_{\odot}-1.4\,M_{\odot}$ NS-NS,
 %$10\,M_{\odot}-1.4\,M_{\odot}$ BH-NS and $10\,M_{\odot}-10\,M_{\odot}$ BH-BH binaries.
 For each of these three configurations, we compute the match between signal and template waveforms for different values of the initial orbital eccentricity $e_0$ between $0$ and $0.4$ (defined at the cut-off frequency $20\,$Hz).
 Fig.~\ref{fig:e02_e06} suggests that the importance of higher-order eccentricity corrections for GW data analysis is strongly dependent on the total mass of an eccentric compact binary source. 
 Given the same $e_0$ but for configurations with increasing total mass, we find that templates restricted to leading-order eccentricity corrections become increasingly faithful representations of those inspiral waveforms that include higher-order eccentricity effects at each PN order. 
 This is expected, as compact binaries with higher total mass provide a smaller number of inspiral GW cycles in the frequency 
window of Advanced LIGO. Therefore, these systems require larger initial eccentricities to bring on a substantial de-phasing and subsequent mismatch between our inspiral signal and template families.
Fig.~\ref{fig:e02_e06} indicates that a waveform model restricted to only leading-order eccentricity corrections would be an effectual template family for the detection of GWs from even moderately eccentric inspirals (with $e_0 \leq 0.15$ and $ \leq 0.3$ for our traditional NS-NS and BH-BH binaries, respectively).
However, the inset of Fig.~\ref{fig:e02_e06} suggests that waveform effects of higher-order eccentricity corrections become distinguishable from detector noise at significantly lower initial eccentricities ($e_0 \geq 0.07$ and $\geq 0.17$ for GWs from NS-NS and BH-BH systems with SNR$\,= 30$). In this region of parameter space, we should expect systematic errors in source parameter estimation with inspiral templates that are accurate only to leading order in eccentricity $e_0$. The inclusion of higher-order eccentricity corrections in waveform modeling is therefore desirable for an accurate follow-up of eccentric GW signals.

We move on to probe data analysis implications of including the effect of periastron advance in our eccentric inspiral waveforms $\tilde{h}(f)$. %, given by Eq.~(\ref{hf_newt}).
In our match calculation $\mathcal{M}(h_s, h_t)$, the signal waveforms employ our quadrupolar-order $\tilde{h}(f)$ given by Eq.~(\ref{hf_newt}), including both $k$ and $e_t$ effects to the sixth order in $e_0$ at each PN order. We build a template family $h_t$ that neglects effects of periastron advance, by extending to 3PN order previously developed eccentric inspiral waveforms (provided with 2PN-accurate Fourier phase in Ref.~\cite{THG}).
In other words, we construct quadrupolar templates $\tilde{h}_t(f)$ with the help of Eq.~(\ref{5}) and 
%, given by our Eqs.~(\ref{5}),(\ref{6.b}),(\ref{6.c}),(\ref{appendixet}),
the 3PN extension of our Newtonian Eq.~(\ref{8}) for $\Psi_j$ 
while  incorporating all 
$\mathcal{O}(e_0^6)$ corrections at each PN order.
Additionally, we evaluate the 
Fourier phase at the unperturbed stationary point $F=f/j$ \cite{YABW}.
It is important to note that such a template waveform family ignores the effect of periastron advance in its Fourier phase evolution.
We consider the same NS-NS, NS-BH and BH-BH systems as before and compute the match between signal and template waveforms for discrete values of initial orbital eccentricity at $20\,$Hz, $e_0 \in [0,0.4]$.
From our results, presented in Fig.~\ref{fig:aop_waop}, we learn that the significance of periastron advance effects for GW data analysis is rather independent of the total mass of the source, with similar match estimates for all three traditional compact binaries under consideration.
Periastron advance starts to influence the effectualness of GW templates for detection only for systems that have eccentricities $e_0 > 0.25$ at $20\,$Hz. This agrees with our observation that $k$-induced modulations in the inspiral waveforms presented in Fig.~5 and 6 of Ref.~\cite{DGI} become clearly visible only for moderate values of initial orbital eccentricity.
However, we can expect systematic biases in the source parameter estimation for much smaller values of orbital eccentricity.
The inset of Fig.~\ref{fig:aop_waop} suggests that periastron advance effects in an eccentric GW signal with SNR$\,=30$ would already become distinguishable from noise for eccentricities $e_0 > 0.03$ at $20\,$Hz, leading to systematic errors in the recovered source parameters when waveform models neglect periastron advance.

Lastly, we explore the relevance of PN-accurate amplitude corrections while constructing realistic analytic Fourier-domain waveforms for eccentric inspirals. For these $\mathcal{M}$ estimates, we invoke as the expected GW signal our 1PN-accurate amplitude corrected $\tilde{h}(f)$, given by Eq.~(\ref{hf_1PN}), 
including the effects of 3PN-accurate periastron advance, frequency and eccentricity evolution accurate to sixth order 
in orbital eccentricity. For the template family, we are utilizing a quadrupolar-order $\tilde{h}(f)$, given by Eq.~(\ref{hf_newt}), that includes the same order effects of 3PN-accurate periastron advance and 3PN-accurate frequency and eccentricity evolution as above.
We consider five compact binary configurations with a fixed total mass $m = m_1 + m_2 = 20 M_\odot$ and varying mass ratios $q = m_1/m_2 \in \{1,3,5,7,9\}$. For each of these configurations, we pursue match computations for different choices of initial orbital eccentricity $e_0 \in [0,0.4]$ at $20\,$Hz, resulting in Fig.~\ref{fig:Newt_1PN}. We observe that amplitude corrections are rather unimportant while constructing template waveforms for equal-mass binaries in eccentric orbits. This is expected, as the dominant amplitude corrections -- appearing at 0.5PN order in Eq.~(\ref{hf_1PN}) -- are proportional to $\sqrt{1-4\eta}$ and therefore vanish for equal-mass binaries. Our plots suggest that the effect of amplitude corrections on the faithfulness of eccentric inspiral waveforms crucially depends on the mass ratio of a binary system, with $\mathcal{M}$ rapidly dropping below the critical value of 0.97 as $q \geq 5$, even for systems with negligible initial eccentricities. This is a familiar result from the modeling of compact binary inspiral along circular orbits and points to the relevance of higher modes for GWs from binaries with asymmetric masses \cite{CVBASG}. In other words, our plots in Fig.~\ref{fig:Newt_1PN} essentially confirm previous literature that compared restricted and amplitude-corrected $\tilde{h}(f)$ for quasi-circular inspiral. Interestingly, we find that the $q$-dependent effect of amplitude corrections on the faithfulness of eccentric inspiral waveforms is largely unaffected by the value of initial eccentricity $e_0$.

\section{Conclusions} \label{conclusion}

We have provided fully analytic PN-accurate Fourier domain gravitational waveforms 
for compact binaries inspiraling along precessing moderately eccentric orbits.
Our inspiral approximant contains 1PN-accurate amplitude corrections and its Fourier phase 
incorporates the effects of 3PN-accurate periastron advance and GW emission. 
Additionally, the eccentricity effects are
accurate to sixth order in $e_0$ at each PN order. 
We infer from our analytic waveform expression that the orbital eccentricity induced 
higher harmonics are no longer integer multiples of orbital frequency due to 
the influence of periastron advance. This substantiates and extends  what is detailed in Ref.~\cite{YS} 
for compact binaries inspiraling along PN-accurate precessing eccentric orbits.
Preliminary GW data analysis implications of our waveforms are probed with the help of the usual match computations.

In what follows, we provide a step-by-step summary of our effort.
\begin{enumerate}
\item

We start from our Eqs.~(\ref{hx0}) and (\ref{hp0})  that provide quadrupolar order GW polarization states from compact
binaries in PN-accurate eccentric orbits as a sum over various harmonics.

\item
With above inputs, we compute the time domain GW detector response function and express it
as a summation of several cosine functions whose arguments are sum of integer multiples of
$\phi $ and $\phi'$ associated with the orbital and periastron motions. Amplitudes of these
functions are expressed in terms of $\omega$, $e_t$ and the angles that specify
the antenna patterns $F_{\times}, F_+ $ and the direction of the orbital
angular momentum vector.  The quadrupolar version of $h(t)$
that  explicitly incorporates the next to leading order $e_t$ corrections
is given by Eq.~(\ref{Eq_h_t_0f}) and associated expressions like
Eqs.~(\ref{gamma_0}) and (\ref{sigma_0}).
 Its 1PN extension is symbolically provided by Eq.~(\ref{ht_1PN})
and the accompanying \texttt{Mathematica} file provide
the explicit expressions for various PN coefficients
while incorporating ${\cal O}(e_t^6)$ corrections.

\item
We also provide a prescription to obtain temporally evolving $h(t)$ for
compact binaries inspiraling due to 3PN accurate GW emission along precessing
3PN accurate orbits of moderate eccentricities.
This involves imposing temporal evolution for $\omega, e_t, \phi'$ and $\phi$
with the help of PN accurate differential equations.
The relative 3PN accurate equations for $\omega$ and $e_t$ are due to the emission of GWs,
as evident from our Eqs.~(\ref{Tevolve_3}) and (\ref{Tevolve_4}).
The conservative 3PN accurate differential equation for
$\phi'$ arises essentially due to periastron advance as evident from Eq.~(\ref{Tevolve_2}).
The differential equation for $\phi$ is kinematical in nature as $d\phi/dt \equiv \omega$.

\item
The structure of the time domain response function allows us to involve the method of stationary phase
approximation to compute its Fourier transform.
The crucial Fourier phases and the associated 'nine'; stationary points may be concisely
written as  $\Psi^{\pm n}(t) \coloneqq -2\pi f t\,+\,j\phi-(j \pm n)\phi'$,
where $n$ takes values $0,1,2,3,4$.
The nine stationary points, associated with 1PN accurate amplitude corrected $h(t)$,
essentially provide relations between the orbital and Fourier frequencies
$ F ( t^{\pm n}) = {f}/ (j-(j \pm n)\, k')$, where $k'$ is related to the rate of periastron
advance per orbit.
The explicit expression for the  resulting 3PN -accurate  Fourier phases with leading order initial eccentricity
corrections are provided by Eq.~(\ref{psi_e02}).
Gathering various results, we obtain Eqs.~(\ref{hf_newt}) and  (\ref{xi_newt})
that provide the quadrupolar order $\tilde{h}(f)$ while incorporating
 fourth order orbital eccentricity contributions along with the effects due to
 3PN-accurate frequency, eccentricity evolution and periastron advance.
Additionally, we have extended these results by including 1PN accurate amplitude
corrections and six order eccentricity contributions.
\item

A crucial ingredient to obtain a fully analytic $\tilde{h}(f)$ involves
a derivation, detailed in Sec.~\ref{sec:PostCirc}, that provides PN accurate analytic expression for
$e_t$ in terms of $e_0, \omega, \omega_0$. We have obtained 3PN accurate expression for $e_t(e_0,\omega,\omega_0)$ by extending the post-circular scheme of Refs.~\cite{YABW,THG}.

\end{enumerate} 

 A number of extensions are possible. 
Influenced by Refs.~\cite{KG05,Kidder95},  we are incorporating the effects of leading order 
aligned spin-orbit and spin-spin interactions into these waveforms. 
 It will be interesting to explore data analysis implications of our present waveforms.
A possible avenue is to explore the astrophysical implications of using PN-accurate periastron 
advance contributions that depend both on $m$ and $\eta$, influenced by 
Refs.~\cite{MKFV,NSeto}.
%by extending 
%what is pursued in Ref.~\cite{MKFV,Abdul10,NSeto}. NOTE: REFERENCE TO NAOKI SETO AND ABDUL H. MROU\'e ADDED. BENCE KOCKSIS WAS ALREADY CITED. }
There are on-going efforts to construct analytic IMR templates to model  eccentric 
compact binary coalescence \cite{Huerta,ENIGMA}. The present waveform family will be relevant to construct IMR templates for moderately eccentric compact binary mergers which can be used to extract orbital eccentricity and periastron advance as done in Ref~\cite{Abdul10}. Efforts are on-going to obtain various constructs, using elements of our post-circular Fourier domain approximant,
that should allow us to make comparisons with brand new PN-accurate frequency domain waveform family, developed in Refs.~\cite{MRLY18,MY19} for moderate eccentricities.

\section{Acknowledgements}
We thank Yannick Boetzel for helpful discussions, suggestions and providing us
the lengthy eccentricity enhancement functions.
%We are grateful to Marc Favata for a critical review of this work. We also thank Blake Moore for carefully reading the manuscript.
We are grateful to Marc Favata  and Blake Moore for their helpful comments. 
M.~H. acknowledges support from Swiss National Science Foundation (SNSF) grant IZCOZ0\_177057. 
We have used software packages from {\tt PyCBC} \cite{alex_nitz_2019_2556644} and {\tt Matplotlib} \cite{matplotlib} to compute and plot match estimates.

\onecolumngrid
\pagebreak

\twocolumngrid
\appendix

\section{$\Gamma^{(0)}_{j,\pm n}$ and $\Sigma^{(0)}_{j,\pm n}$ coefficients} \label{appendixB} 
We list the $\Gamma^{(0)}_{j,\pm n}$ and $\Sigma^{(0)}_{j,\pm n}$ coefficients appearing in Eq.(\ref{hpx0_1}). The relevant $\Gamma_{j,\pm n}^{(0)}$ 
expressions read 
\begin{widetext}
\begin{subequations} \label{gamma_0}
\begin{align}
\Gamma_{1,-2}^{(0)} &=F_+ \bigg\{\bigg(\frac{3 e_t}{2}-\frac{13 e_t^3}{16}\bigg) \left(1+c_i^2\right) c_{2 \beta }\bigg\}+F_\times \bigg\{\bigg(-3 e_t+\frac{13 e_t^3}{8}\bigg) c_i s_{2 \beta }\bigg\}, \\
\Gamma_{2,-2}^{(0)} &=F_+ \bigg\{\bigg(-2+5 e_t^2-\frac{23 e_t^4}{8}\bigg) \left(1+c_i^2\right) c_{2 \beta }\bigg\}+F_\times \bigg\{\bigg(4-10 e_t^2 +\frac{23 e_t^4}{4}\bigg) c_i s_{2 \beta }\bigg\}, \\
\Gamma_{3,-2}^{(0)} &=F_+ \bigg\{\bigg(-\frac{9 e_t}{2}+\frac{171 e_t^3}{16}\bigg) \left(1+c_i^2\right) c_{2 \beta }\bigg\}+F_\times \bigg\{\bigg(9 e_t-\frac{171 e_t^3}{8}\bigg) c_i s_{2 \beta }\bigg\}, \\
\Gamma_{4,-2}^{(0)} &=F_+ \bigg\{\left(-8 e_t^2+20 e_t^4\right) \left(1+c_i^2\right) c_{2 \beta }\bigg\}+F_\times \bigg\{\bigg(16 e_t^2-40 e_t^4\bigg) c_i s_{2 \beta }\bigg\}, \\
\Gamma_{5,-2}^{(0)} &=F_+ \bigg\{-\frac{625}{48} e_t^3 \left(1+c_i^2\right) c_{2 \beta }\bigg\}+F_\times \bigg\{\frac{625}{24} e_t^3 c_i s_{2 \beta }\bigg\}, \\ 
\Gamma_{6,-2}^{(0)} &=F_+ \bigg\{-\frac{81}{4} e_t^4 \left(1+c_i^2\right) c_{2 \beta }\bigg\}+F_\times \bigg\{\frac{81}{2} e_t^4 c_i s_{2 \beta }\bigg\}, \\
\Gamma_{1,0}^{(0)} &=F_+ \bigg\{\bigg(e_t-\frac{e_t^3}{8}\bigg) s_i^2\bigg\}, \\
\Gamma_{2,0}^{(0)} &=F_+ \bigg\{\bigg(e_t^2-\frac{e_t^4}{3}\bigg) s_i^2\bigg\}, \\
\Gamma_{3,0}^{(0)} &=F_+ \bigg\{\frac{9}{8} e_t^3 s_i^2\bigg\}, \\
\Gamma_{4,0}^{(0)} &=F_+ \bigg\{\frac{4}{3} e_t^4 s_i^2\bigg\}, \\
\Gamma_{1,+2}^{(0)} &=F_+ \bigg\{\frac{7}{48} e_t^3 \left(1+c_i^2\right) c_{2 \beta }\bigg\}+F_\times \bigg\{-\frac{7}{24} e_t^3 c_i s_{2 \beta }\bigg\}, \\
\Gamma_{2,+2}^{(0)} &=F_+ \bigg\{-\frac{1}{8} e_t^4 \left(1+c_i^2\right) c_{2 \beta }\bigg\}+F_\times \bigg\{-\frac{1}{4} e_t^4 c_i s_{2 \beta }\bigg\}\,.
\end{align}
\end{subequations}
The $ \Sigma_{j,\pm n}^{(0)}$ counterparts of above expressions read 
\begin{subequations} \label{sigma_0}
\begin{align}
\Sigma_{1,-2}^{(0)} &=F_+ \bigg\{\bigg(\frac{3 e_t}{2}-\frac{13 e_t^3}{16}\bigg) \left(1+c_i^2\right) s_{2 \beta }\bigg\}+F_\times \bigg\{\bigg(3 e_t-\frac{13 e_t^3}{8}\bigg) c_i c_{2 \beta }\bigg\}, \\
\Sigma_{2,-2}^{(0)} &=F_+ \bigg\{\bigg(-2+5 e_t^2-\frac{23 e_t^4}{8}\bigg) \left(1+c_i^2\right) s_{2 \beta }\bigg\}+F_\times \bigg\{\bigg(-4+10 e_t^2-\frac{23 e_t^4}{4}\bigg) c_i c_{2 \beta }\bigg\}, \\
\Sigma_{3,-2}^{(0)} &=F_+ \bigg\{\bigg(-\frac{9 e_t}{2}+\frac{171 e_t^3}{16}\bigg) \left(1+c_i^2\right) s_{2 \beta }\bigg\}+F_\times \bigg\{\bigg(-9 e_t+\frac{171 e_t^3}{8}\bigg) c_i c_{2 \beta }\bigg\}, \\
\Sigma_{4,-2}^{(0)} &=F_+ \bigg\{\left(-8 e_t^2+20 e_t^4\right) \left(1+c_i^2\right) s_{2 \beta }\bigg\}+F_\times \bigg\{\bigg(-16 e_t^2 +40 e_t^4\bigg) c_i c_{2 \beta }\bigg\}, \\
\Sigma_{5,-2}^{(0)} &=F_+ \bigg\{-\frac{625}{48} e_t^3 \left(1+c_i^2\right) s_{2 \beta }\bigg\}+F_\times \bigg\{-\frac{625}{24}  e_t^3 c_i c_{2 \beta }\bigg\}, \\
\Sigma_{6,-2}^{(0)} &=F_+ \bigg\{-\frac{81}{4} e_t^4 \left(1+c_i^2\right) s_{2 \beta }\bigg\}+F_\times \bigg\{-\frac{81}{2} e_t^4 c_i c_{2 \beta }\bigg\}, \\
\Sigma_{1,0}^{(0)} &=0, \\
\Sigma_{2,0}^{(0)} &=0, \\
\Sigma_{3,0}^{(0)} &=0, \\
\Sigma_{4,0}^{(0)} &=0, \\
\Sigma_{1,+2}^{(0)} &=F_+ \bigg\{-\frac{7}{48} e_t^3 \left(1+c_i^2\right) s_{2 \beta }\bigg\}+F_\times \bigg\{-\frac{7}{24} e_t^3 c_i c_{2 \beta }\bigg\}, \\
\Sigma_{2,+2}^{(0)} &=F_+ \bigg\{-\frac{1}{8} e_t^4 \left(1+c_i^2\right) s_{2 \beta }\bigg\}+F_\times \bigg\{-\frac{1}{4} e_t^4 c_i c_{2 \beta }\bigg\}.
\end{align}
\end{subequations}
\end{widetext}

\section{3PN accurate $\frac{d\omega}{dt}$ and $\frac{de_t}{dt}$} \label{appendixC} 

We give here the 3PN accurate expressions for temporal evolution of $\omega$ and $e_t$ for obtaining $h(t)$ associated with compact binaries inspiraling along precessing eccentric orbits. 1PN accurate $\frac{d\omega}{dt}$ and $\frac{de_t}{dt}$ with $\mathcal{O}(e_t^6)$ eccentricity corrections are given by Eq.(\ref{Tevolve_3}) and Eq.(\ref{Tevolve_4}) respectively. The 1.5PN - 3PN contributions to $\frac{d\omega}{dt}$ appearing in Eq. \ref{Tevolve_3} with $\mathcal{O}(e_t^6)$ corrections are,
\begin{widetext}
\begin{subequations}
\begin{align}
\dot{\omega}^{\tiny {1.5PN}} = & \,\pi\,x^{3/2}\bigg\{4+\frac{2335}{48}e_t^2+\frac{42955}{192}e_t^4+\frac{6204647}{9216}e_t^6\bigg\}, \\ \nonumber \\
\dot{\omega}^{\tiny {2PN}} = & \, x^2\bigg\{\frac{34103}{18144}+\frac{13661}{2016}\eta+\frac{59}{18}\eta^2 +\bigg(-\frac{479959}{12096}+\frac{80425}{4032}\eta+\frac{213539}{1728}\eta^2\bigg)e_t^2+\bigg(-\frac{2932261}{16128}-\frac{5715083}{16128}\eta \nonumber \\ & +\frac{2133235}{2304}\eta^2\bigg)e_t^4 +\bigg(-\frac{19581787}{48384} -\frac{1753627}{768}\eta+\frac{25727065}{6912}\eta^2\bigg)e_t^6\bigg\},  \\ \nonumber \\
\dot{\omega}^{\tiny {2.5PN}} = & \,\pi\,x^{5/2}\bigg\{-\frac{4159}{672}-\frac{189}{8}\eta+\bigg(\frac{7885}{96}-\frac{27645}{56}\eta\bigg) e_t^2+\bigg(\frac{44644883}{43008}-\frac{11707809}{3584}\eta\bigg) e_t^4 \nonumber \\ & +\bigg(\frac{971752501}{193536}-\frac{103819241}{8064}\eta\bigg) e_t^6\bigg\}, \\ \nonumber \\
\dot{\omega}^{\tiny {3PN}} = & \, x^3\bigg\{\frac{16447322263}{139708800}+\frac{16 \pi ^2}{3}-\frac{1712 \gamma }{105}+\bigg(-\frac{56198689}{217728} +\frac{451 \pi ^2}{48}\bigg) \eta +\frac{541}{896}\eta ^2-\frac{5605}{2592}\eta ^3-\frac{3424 \log (2)}{105}\nonumber \\ &-\frac{856 \log (x)}{105}+\bigg(\frac{277391496167}{139708800}+\frac{992 \pi^2}{9}-\frac{106144 \gamma }{315} +\bigg(-\frac{280153957}{120960}+\frac{188231 \pi ^2}{2304}\bigg) \eta -\frac{73109}{448}\eta ^2 \nonumber \\ &-\frac{6874115}{31104}\eta^3 -\frac{80464 \log (2)}{315}-\frac{234009 \log (3)}{560}-\frac{53072 \log (x)}{315}\bigg) e_t^2  +\bigg(\frac{974308007423}{79833600}+\frac{3059\pi ^2}{4}-\frac{46759 \gamma }{20}\nonumber \\ &+\bigg(-\frac{33126017}{3780}+\frac{2065129 \pi ^2}{6144}\bigg) \eta  -\frac{2804209}{32256}\eta ^2-\frac{114255295}{41472}\eta ^3  -\frac{2730533 \log (2)}{252}+\frac{4446171 \log (3)}{2240}\nonumber \\ &-\frac{46759 \log (x)}{40}\bigg)e_t^4+\bigg(\frac{150878591021}{3193344}+\frac{76615 \pi ^2}{24}-\frac{234223 \gamma }{24} +\bigg(-\frac{7739324653}{362880}+\frac{34978699 \pi^2}{36864}\bigg) \eta \nonumber \\ & +\frac{21116263}{4608}\eta ^2-\frac{1935750565}{124416}\eta ^3+\frac{80906873 \log (2)}{2520}-\frac{134711181 \log(3)}{35840}-\frac{5224609375 \log (5)}{193536} \nonumber \\ & -\frac{234223 \log (x)}{48}\bigg) e_t^6\bigg\},
\, \label{Tevolve_5} \\ \nonumber \\ \nonumber
\end{align}
\end{subequations}
\end{widetext}
where $\gamma$ stands for the Euler-Mascheroni constant. The 1.5PN - 3PN contributions to $\frac{de_t}{dt}$ appearing in Eq. \ref{Tevolve_4} with $\mathcal{O}(e_t^6)$ corrections are,
\begin{widetext}
\begin{subequations}
\begin{align}
\dot{e_t}^{\tiny {1.5PN}} = & \, \pi\,x^{3/2}\bigg\{\frac{985}{152}+\frac{21729}{608}e_t^2+\frac{3061465}{29184}e_t^4+\frac{161865935}{700416}e_t^6\bigg\}, \\ \nonumber \\
\dot{e_t}^{\tiny {2PN}} = & \,x^2\bigg\{-\frac{108197}{38304}+\frac{56407}{4256}\eta +\frac{141}{19}\eta ^2+\bigg(-\frac{1368625}{51072}-\frac{288209}{17024}\eta+\frac{274515}{2432}\eta^2\bigg) e_t^2 +\bigg(-\frac{15037865}{306432} \nonumber \\ & -\frac{30369109}{102144}\eta+\frac{7578425}{14592}\eta ^2\bigg)e_t^4  +\bigg(-\frac{13488023}{408576}-\frac{65394101}{58368}\eta+\frac{87633595}{58368}\eta ^2\bigg) e_t^6\bigg\}, \\ \nonumber \\ 
\dot{e_t}^{\tiny {2.5PN}} = & \,\pi\,x^{5/2}\bigg\{-\frac{55691}{4256}  -\frac{19067}{399}\eta+\bigg(\frac{286789}{3584}-\frac{7810371}{17024}\eta\bigg)e_t^2  +\bigg(\frac{535570255}{817152}-\frac{31241795}{16128}\eta\bigg) e_t^4 \\ \nonumber &+\bigg(\frac{92235604259}{39223296}-\frac{164170915723}{29417472}\eta\bigg) e_t^6\bigg\}, \\ \nonumber \\
\dot{e_t}^{\tiny {3PN}} =& \,x^3\bigg\{\frac{246060953209}{884822400}+\frac{769 \pi ^2}{57} -\frac{82283 \gamma }{1995}+\bigg(-\frac{613139897}{2298240}+\frac{22345 \pi ^2}{3648}\bigg) \eta-\frac{1046329}{51072}\eta ^2-\frac{305005}{49248}\eta ^3 \nonumber \\ & -\frac{11021 \log (2)}{285}-\frac{234009 \log (3)}{5320}-\frac{82283 \log(x)}{3990}+\bigg(\frac{1316189396351}{589881600}+\frac{14023 \pi ^2}{114}-\frac{1500461 \gamma }{3990} \nonumber \\ & +\bigg(-\frac{5882746699}{4596480} +\frac{46453 \pi^2}{1536}\bigg) \eta -\frac{554719}{4788}\eta ^2-\frac{100330729}{393984}\eta ^3-\frac{2271503 \log (2)}{1330}+\frac{6318243 \log(3)}{21280}\nonumber \\ &-\frac{1500461 \log (x)}{7980}\bigg) e_t^2  +\bigg(\frac{1499268531223}{168537600}+\frac{10129 \pi ^2}{19}-\frac{154829 \gamma}{95}+\bigg(-\frac{543123237}{170240}+\frac{2360575 \pi ^2}{29184}\bigg) \eta \nonumber \\ & +\frac{36456205}{87552}\eta ^2  -\frac{1523467085}{787968}\eta^3 +\frac{41683669 \log (2)}{5985}-\frac{281044809 \log (3)}{340480}-\frac{1044921875 \log (5)}{204288} \nonumber \\ &-\frac{154829 \log (x)}{190}\bigg)e_t^4  +\bigg(\frac{682257052877}{26966016}+\frac{976185 \pi ^2}{608}-\frac{2984337 \gamma }{608}+\bigg(-\frac{4722976831}{875520}+\frac{24558057 \pi^2}{155648}\bigg) \eta  \nonumber \\ &+\frac{1312493803}{350208}\eta ^2  -\frac{24620050735}{3151872}\eta ^3 -\frac{10971071339 \log (2)}{191520}-\frac{74286859077\log (3)}{2723840}+\frac{24033203125 \log (5)}{700416} \nonumber \\ &-\frac{2984337 \log (x)}{1216}\bigg) e_t^6\bigg\}. \label{Tevolve_6} \\ \nonumber
\end{align}
\end{subequations}
\end{widetext}

\section{3PN accurate analytic expressions for $e_t$ and $\Psi_j^{\pm n}$ } \label{appendixA} 

 We display explicit expressions for 3PN-accurate $e_t$ and Fourier phases
 that incorporate next to leading order $e_0$ corrections at each PN order.
 These expressions along with Eqs. (\ref{hf_newt}), (\ref{xi_newt}), (\ref{alpha_phibr}), (\ref{gamma_0}) and (\ref{sigma_0}) are 
 % 3.13,3.14, 2.33, 2.22 and 2.23
 required to make operational the fully analytic frequency domain 
 quadrupolar order GW response function for eccentric inspirals
 that includes ${\cal O}(e_0^4)$ corrections at every PN order.
 We begin by listing explicit expression for the 3PN accurate 
 $e_t$ in terms of $e_0,\chi$ and $x$. The underlying computation is detailed
 in Ref.~\cite{THG} and requires 3PN-accurate expressions for 
 $\dot \omega$ and $\dot e_t$, given by Eqs.~(\ref{Tevolve_3}) and (\ref{Tevolve_4}). The fully 3PN accurate $e_t$ expression that accounts 
 for all the ${\cal O}(e_0^3)$ contributions read 
%\begin{widetext}
\begin{align}                           \label{appendixet}
e_t  = \sum_{m=0}^{6} \mathcal{E}_m x^{m/2}\,.    
\end{align}
The coefficients  $\mathcal{E}_m$ with next to leading order eccentricity corrections ${\cal O}{\left(e_0^3\right)}$ at each PN order can be listed as,
\begin{widetext}
\begin{subequations}
\begin{align}
&\mathcal{E}_0=e_0 \chi ^{-19/18} + \frac{3323}{1824} \bigg(\chi ^{-19/18}-\chi ^{-19/6}\bigg) e_0^3, \\ \nonumber \\
&\mathcal{E}_1=0, \\ \nonumber \\
&\mathcal{E}_2=\bigg\{\bigg(-\frac{2833}{2016}+\frac{197 \eta }{72}\bigg)\chi^{-19/18}+\bigg(\frac{2833}{2016}-\frac{197\eta}{72}\bigg)\chi^{-31/18}\bigg\}e_0 + \bigg\{\bigg(-\frac{9414059}{3677184}+\frac{654631 \eta }{131328}\bigg)\chi^{-19/18} \nonumber \\ &\qquad+\bigg(\frac{386822573}{47803392}-\frac{1482433 \eta }{131328}\bigg)\chi^{-31/18}+\bigg(\frac{11412055}{5311488}-\frac{378697\eta}{43776}\bigg)\chi^{-19/6} \nonumber \\ &\qquad+\bigg(-\frac{9414059}{1225728}+\frac{654631 \eta }{43776}\bigg) \chi ^{-23/6} \bigg\}e_0^3, \\ \nonumber \\
&\mathcal{E}_3=\bigg\{\frac{377}{144}\pi\bigg(-\chi^{-19/18}+\chi ^{-37/18}\bigg)\bigg\}e_0 + \bigg\{-\frac{1252771 \pi }{262656} \chi ^{-19/18}+\frac{1315151 \pi}{131328} \chi ^{-37/18}+\frac{396797 \pi}{43776}\chi ^{-19/6} \nonumber \\ &\qquad-\frac{1252771 \pi}{87552}  \chi^{-25/6} \bigg\}e_0^3 , \\ \nonumber \\
&\mathcal{E}_4=\bigg\{\bigg(\frac{77006005}{24385536}-\frac{1143767 \eta }{145152}+\frac{43807 \eta ^2}{10368}\bigg)\chi^{-19/18}+\bigg(-\frac{8025889}{4064256}+\frac{558101\eta }{72576}-\frac{38809 \eta^2}{5184}\bigg)\chi^{-31/18} \nonumber \\ & \qquad +\bigg(-\frac{28850671}{24385536}+\frac{27565 \eta }{145152}+\frac{33811 \eta^2}{10368}\bigg) \chi ^{-43/18}\bigg\}e_0+\bigg\{\bigg(\frac{255890954615}{44479217664}-\frac{3800737741 \eta }{264757248} \nonumber \\ & \qquad +\frac{145570661 \eta ^2}{18911232}\bigg) \chi^{-19/18}+\bigg(-\frac{1095868349309}{96371638272}+\frac{65400285919 \eta }{1720922112}-\frac{292039301 \eta ^2}{9455616}\bigg) \chi^{-31/18} \nonumber \\ & \qquad +\bigg(-\frac{20952382669619}{4047608807424}-\frac{385200824731 \eta }{24092909568}+\frac{4301644427 \eta ^2}{132378624}\bigg) \chi^{-43/18}+\bigg(\frac{8180980796033}{1349202935808}\nonumber \\ & \qquad+\frac{14604819923 \eta }{2676989952}-\frac{317361763 \eta ^2}{14708736}\bigg) \chi^{-19/6}+\bigg(\frac{32330351815}{3569319936} -\frac{10345778159 \eta }{191213568} +\frac{74603309 \eta ^2}{1050624}\bigg) \chi^{-23/6} \nonumber \\ & \qquad +\bigg(-\frac{9164199307}{2118057984}+\frac{1205846917 \eta }{29417472}-\frac{13714021 \eta ^2}{233472}\bigg) \chi ^{-9/2}\bigg\}e_0^3, \\ \nonumber \\
&\mathcal{E}_5=\bigg\{\bigg(\frac{9901567 \pi }{1451520}-\frac{202589 \pi  \eta }{362880}\bigg) \chi ^{-19/18}+\bigg(-\frac{1068041 \pi }{290304}+\frac{74269 \pi  \eta}{10368}\bigg) \chi ^{-31/18}+\bigg(-\frac{1068041 \pi }{290304} \nonumber \\ & \qquad+\frac{74269 \pi  \eta }{10368}\bigg) \chi ^{-37/18}+\bigg(\frac{778843 \pi}{1451520}-\frac{4996241 \pi  \eta }{362880}\bigg) \chi ^{-49/18}\bigg\}e_0+\bigg\{\bigg(\frac{32902907141 \pi }{2647572480}\nonumber \\ &\qquad-\frac{673203247 \pi  \eta }{661893120}\bigg) \chi ^{-19/18}  +\bigg(-\frac{11217854617 \pi}{529514496}+\frac{558877241 \pi  \eta }{18911232}\bigg) \chi ^{-31/18}+\bigg(-\frac{3725822783 \pi }{264757248}\nonumber \\ & \qquad+\frac{259084747 \pi  \eta}{9455616}\bigg) \chi ^{-37/18}+\bigg(\frac{195499289159 \pi }{2647572480}-\frac{65776041763 \pi  \eta }{661893120}\bigg) \chi^{-49/18}+\bigg(-\frac{2057616403 \pi }{32686080}\nonumber \\ & \qquad+\frac{2370731599 \pi  \eta }{73543680}\bigg) \chi ^{-19/6}+\bigg(\frac{1124125901 \pi}{29417472}-\frac{78169009 \pi  \eta }{1050624}\bigg) \chi ^{-23/6}+\bigg(\frac{330949595 \pi }{19611648}\nonumber \\ & \qquad-\frac{142768769 \pi  \eta }{2101248}\bigg)\chi ^{-25/6}+\bigg(-\frac{12693032573 \pi }{294174720}  +\frac{11292740311 \pi  \eta }{73543680}\bigg) \chi ^{-29/6}\bigg\}e_0^3 \\ \nonumber 
\end{align}
\end{subequations}
\end{widetext}
Due to the lengthy nature of 3PN order terms in $e_t$, we split it in two parts 
as 
%We rewrite the coefficient of $x^3$ term in $e_t$ due to its expanse as, 
\begin{align}
\mathcal{E}_6=\mathcal{E}_6^{'}e_0 + \mathcal{E}_6^{''}e_0^3,  \\ \nonumber 
\end{align}
%with coefficients of leading order as well as next to leading order eccentricity correctios at 3PN as,
The explicit form of these two contributions are 
\begin{widetext}
\begin{subequations}
\begin{align}
&\mathcal{E}_6^{'}= \bigg(-\frac{33320661414619}{386266890240}+\frac{180721 \pi^2}{41472}+\frac{3317 \gamma }{252}+\bigg(\frac{161339510737}{8778792960}+\frac{3977 \pi ^2}{2304}\bigg) \eta -\frac{359037739 \eta^2}{20901888} +\frac{10647791 \eta ^3}{2239488} \nonumber \\ & \qquad+\frac{12091 \log (2)}{3780}+\frac{26001 \log (3)}{1120}+\frac{3317 \log (x)}{504}\bigg)\chi ^{-19/18}+\bigg(\frac{218158012165}{49161240576}-\frac{34611934451 \eta }{1755758592}+\frac{191583143 \eta ^2}{6967296} \nonumber \\ & \qquad-\frac{8629979 \eta ^3}{746496}\bigg) \chi^{-31/18}-\frac{142129 \pi ^2}{20736}\chi ^{-37/18} +\bigg(\frac{81733950943}{49161240576}-\frac{6152132057 \eta }{1755758592}-\frac{1348031 \eta^2}{331776} \nonumber \\ & \qquad+\frac{6660767 \eta ^3}{746496}\bigg) \chi ^{-43/18}+\bigg(\frac{216750571931393}{2703868231680}+\frac{103537 \pi ^2}{41472}-\frac{3317 \gamma }{252} +\bigg(\frac{866955547}{179159040}-\frac{3977 \pi^2}{2304}\bigg) \eta \nonumber \\ & \qquad-\frac{130785737 \eta ^2}{20901888}-\frac{4740155 \eta ^3}{2239488}-\frac{12091 \log (2)}{3780}-\frac{26001 \log(3)}{1120} -\frac{3317 \log (x)}{504}-\frac{3317 \log (\chi )}{756}\bigg)\chi ^{-55/18} ,  \\  \nonumber
&\mathcal{E}_6^{''}=\bigg(-\frac{110724557880778937}{704550807797760} +\frac{600535883 \pi^2}{75644928}+\frac{11022391 \gamma }{459648}+\bigg(\frac{536131194179051}{16012518359040}+\frac{13215571 \pi ^2}{4202496}\bigg) \eta \nonumber \\  & \qquad -\frac{1193082406697 \eta ^2}{38125043712}+\frac{35382609493 \eta ^3}{4084826112}+\frac{40178393 \log (2)}{6894720}+\frac{28800441 \log(3)}{680960}+\frac{11022391 \log (x)}{919296}\bigg)\chi ^{-19/18} \nonumber \\  & \qquad +\bigg(\frac{29787660990550865}{1165711336538112}-\frac{591234360321013 \eta }{5947506819072}+\frac{107636760191 \eta ^2}{874119168}-\frac{64940942431 \eta^3}{1361608704}\bigg) \chi ^{-31/18}\nonumber \\ & \qquad-\frac{495811927 \pi^2}{18911232}\chi^{-37/18}+\bigg(\frac{59358100103030627}{8159979355766784}+\frac{2420024232862595 \eta }{291427834134528}-\frac{103398129181999 \eta^2}{1156459659264}\nonumber \\ & \qquad+\frac{847423952119 \eta ^3}{9531260928}\bigg) \chi ^{-43/18}+ \bigg(-\frac{3881667007528080426037}{2243994322835865600} +\frac{720177509 \pi^2}{75644928}+\frac{517414657 \gamma }{2298240} \nonumber \\  & \qquad +\bigg(-\frac{1395931720786001359}{1457139170672640}+\frac{295851449 \pi ^2}{4202496}\bigg)\eta-\frac{112681906698415 \eta ^2}{3469378977792} -\frac{1549239851389 \eta ^3}{28593782784}\nonumber \\ & \qquad+\frac{101727523747 \log (2)}{6894720}-\frac{5477465997 \log(3)}{680960}+\frac{517414657 \log (x)}{4596480}-\frac{517414657 \log (\chi )}{6894720}\bigg)\chi ^{-55/18}\nonumber \\ & \qquad+ \bigg(\frac{152896024020300184249}{67999827964723200}-\frac{95207357\pi ^2}{8404992}-\frac{245954159 \gamma }{766080}+\bigg(\frac{12374839994637661}{10793623486464}-\frac{116237911 \pi ^2}{1400832}\bigg) \eta \nonumber \\ & \qquad-\frac{3908281091711 \eta ^2}{128495517696}-\frac{42680326813 \eta ^3}{1059028992}-\frac{33962745773 \log (2)}{2298240}+\frac{5362264233 \log(3)}{680960}-\frac{245954159 \log (x)}{1532160}\bigg)\chi ^{-19/6}\nonumber \\ & \qquad+\bigg(\frac{23176718595161489}{906664372862976}-\frac{866895029665039 \eta }{32380870459392}-\frac{5814138473063 \eta ^2}{42831839232}+\frac{62520267311 \eta ^3}{353009664}\bigg) \chi ^{-23/6}\nonumber \\ & \qquad+\frac{149592469 \pi^2}{2101248} \chi^{-25/6}+ \bigg(-\frac{99813874374700537}{234850269265920}-\frac{429547595 \pi^2}{8404992}+\frac{11022391 \gamma }{153216}+\bigg(-\frac{62659748948903}{1779168706560}\nonumber \\ & \qquad+\frac{13215571 \pi ^2}{1400832}\bigg) \eta -\frac{95613034561\eta ^2}{1412038656}+\frac{22151672941 \eta ^3}{151289856}+\frac{40178393 \log (2)}{2298240} +\frac{86401323 \log (3)}{680960}+\frac{11022391 \log(x)}{306432} \nonumber \\ & \qquad-\frac{11022391 \log (\chi )}{459648}\bigg)\chi ^{-31/6}+\bigg(\frac{31472267987495}{6167784849408}-\frac{318662569276073 \eta }{4625838637056}+\frac{4844584781833 \eta^2}{18356502528} \nonumber \\ & \qquad-\frac{1562882519 \eta ^3}{5603328}\bigg) \chi ^{-9/2}.  
\end{align}
\end{subequations}
\end{widetext}
We have pursued careful checking of our results with what is available in Ref.~\cite{THG} and observed a slight typo in the $\mathcal{O}(e_0^5)$ contributions for the $e_t$ expression (Eq.~(A6e)) of Ref.~\cite{THG}. The $\eta$ independent term present in  the coefficient of $\chi^{-119/18}$ should be $16952610560003855/162260186038272$ instead of $16633441088056655/162260186038272$. 
Note that the above $e_t$ expression is required while computing the Fourier amplitudes $\xi_j$. Additionally, it is a crucial ingredient while 
computing analytic expression for our Fourier phases $\Psi_j$. 
It should be obvious that its frequency dependence is encapsulated in $\chi=F/F_0$ and the PN expansion parameter $x=\left(G\,m\,2\,\pi\,F/c^3\right)^{2/3}$. 
%Caution is recommended while employing our $e_t$ while 
%evaluating the amplitudes of $\tilde{h}(f)$. We evaluate $\tilde{h}(f)$ 
% at GW frequencies $f$ which are related to the orbital frequencies $F$ 
 %Eq.~() for various harmonics in the 
 %approach of {SPA}.
 %Further, care should be taken while mapping $e_t$ value to the 
 %appropriate $F$ value for a given harmonic. 
 %We let $F=f/j$ and $F_0=f_0/j$ 
 %while evaluating $\chi$, $x$ and $\Psi_j$ 
 %expressions.
 %in amplitude estimates.
 %The above choice for $F_0$ is influenced by the fact that 
 %the third harmonic 
 %being the dominant one for a moderately eccentric system is aptly chosen to describe $e_0$ \cite{MG07}.
 %This allows us to define 
 %$e_0$ to be the eccentricity of the orbit whose {\it third} harmonic first enters the GW detector band at $f_0$.

%It should be noted that the value for $x$ parameter has to be %replaced at the appropriate stationary point as discussed in %Sec.~\ref{sec:level2}.
%\\

  We now display our 3PN accurate closed form expression for the Fourier phases $\Psi_j^{\pm n}$. 
  Recall that  \textit{nine} different Fourier phases appear in our
  1PN accurate amplitude corrected 
  $\tilde{h}(f)$ expression, given by Eq.~(\ref{hf_1PN}). To circumvent the task of displaying all the 9 different Fourier phases separately, we provide a general expression for these phases as $\Psi_j^{n}$ where $n=0,1,2,3,4$. 
  It is not very difficult to obtain $\Psi_j^{\pm n}$ from $\Psi_j^{n}$ 
  by replacing $n$ with appropriate $'\pm'$ sign in the expression. The general expression for 3PN accurate Fourier phase reads 
\begin{widetext}
\begin{align}                \label{appendixpsi}
  \Psi_j^n & = 
    \left(j-(j+n)k^{(6)}_{(3)}\right) \phi_c - 2\pi f t_c - \frac{3 \, j}{256 \, \eta \, x^{5/2}}\sum_{m=0}^{6} \mathcal{P}_m x^{m/2}\,.               
\end{align}
\end{widetext}
Various PN coefficients $\mathcal{P}_m$ with next to leading order eccentricity contributions are given by 
\begin{widetext}
\begin{subequations}
\begin{align}
&\mathcal{P}_0=1-\frac{2355}{1462} e_0^2 \chi ^{-19/9} +\bigg(-\frac{2608555}{444448}\chi ^{-19/9}+\frac{5222765}{998944} \chi ^{-38/9}\bigg) e_0^4, \\ \nonumber \\
&\mathcal{P}_1=0, \\ \nonumber \\
&\mathcal{P}_2=-\frac{2585}{756}-\frac{25 n}{3 j}+\frac{55 \eta }{9}+\bigg\{\bigg(\frac{69114725}{14968128}+\frac{1805 n}{172 j}-\frac{128365 \eta }{12432}\bigg)\chi^{-19/9}+\bigg(-\frac{2223905}{491232}+\frac{154645 \eta }{17544}\bigg) \chi ^{-25/9}\bigg\}
   e_0^2 \nonumber \\ &\qquad +\bigg\{\bigg(\frac{229668231175}{13650932736}+\frac{315685 n}{8256 j}-\frac{426556895 \eta }{11337984}\bigg) \chi^{-19/9}+\bigg(-\frac{14275935425}{416003328}+\frac{209699405 \eta }{4000032}\bigg) \chi
   ^{-25/9} \nonumber \\ &\qquad +\bigg(-\frac{259509826776175}{13976341456896}-\frac{225548425 n}{6014496 j}+\frac{1222893635 \eta }{28804608}\bigg) \chi^{-38/9} \nonumber \\ &\qquad +\bigg(\frac{14796093245}{503467776}-\frac{1028884705 \eta }{17980992}\bigg) \chi ^{-44/9}\bigg\} e_0^4, \\ \nonumber \\
&\mathcal{P}_3=-16 \pi +\bigg(\frac{65561 \pi}{4080}\chi ^{-19/9}-\frac{295945 \pi}{35088}\chi^{-28/9}\bigg) e_0^2+\bigg(\frac{217859203 \pi}{3720960}\chi^{-19/9}-\frac{3048212305 \pi}{64000512}\chi ^{-28/9} \nonumber \\ &\qquad-\frac{6211173025 \pi}{102085632}\chi ^{-38/9}+\frac{1968982405\pi}{35961984}\chi^{-47/9}\bigg) e_0^4, \\ \nonumber \\
&\mathcal{P}_4=-\frac{48825515}{508032}-\frac{31805 n}{252 j}+\left(\frac{22105}{504}-\frac{10 n}{j}\right) \eta +\frac{3085 \eta ^2}{72}+\bigg\{\bigg(\frac{115250777195}{2045440512}+\frac{323580365 n}{5040288 j}+\bigg(-\frac{72324815665}{6562454976}\nonumber \\ &\qquad+\frac{36539875 n}{1260072 j}\bigg) \eta-\frac{10688155 \eta ^2}{294624}\bigg) \chi ^{-19/9}+\bigg(\frac{195802015925}{15087873024}+\frac{5113565 n}{173376 j}+\bigg(-\frac{3656612095}{67356576}-\frac{355585 n}{6192 j}\bigg) \eta\nonumber \\ &\qquad +\frac{25287905 \eta ^2}{447552}\bigg) \chi^{-25/9}+\bigg(\frac{936702035}{1485485568}+\frac{3062285 \eta }{260064}-\frac{14251675 \eta ^2}{631584}\bigg) \chi ^{-31/9}\bigg\}e_0^2 + \bigg\{\bigg(\frac{382978332618985}{1865441746944}\nonumber \\ &\qquad+\frac{1075257552895 n}{4596742656 j}+\bigg(-\frac{240335362454795}{5984958938112}+\frac{121422004625 n}{1149185664 j}\bigg)\eta -\frac{35516739065 \eta ^2}{268697088}\bigg) \chi ^{-19/9}\nonumber \\ &\qquad+\bigg(\frac{1256913822951125}{12777273040896}+\frac{1727660975 n}{7727616 j}+\bigg(-\frac{1182697961961875}{3194318260224}-\frac{25377635 n}{74304 j}\bigg) \eta +\frac{34290527545 \eta ^2}{102041856}\bigg) \chi^{-25/9}\nonumber \\ &\qquad+\bigg(-\frac{94372278903235}{7251965779968}+\frac{126823556396665 \eta }{733829870592}-\frac{20940952805 \eta ^2}{93768192}\bigg) \chi^{-31/9}+\bigg(-\frac{359074780345285439107}{1705190973672775680}\nonumber \\ &\qquad-\frac{100456187745548465 n}{451108723193856 j}+\bigg(-\frac{41964795442387913}{5074973135930880}-\frac{656130734149165 n}{3717929037312 j}\bigg) \eta +\frac{203366083643 \eta
   ^2}{1130734080}\bigg) \chi ^{-38/9}\nonumber \\ &\qquad+\bigg(-\frac{735191339256903775}{7044076094275584}-\frac{638978688025 n}{3031305984 j}+\bigg(\frac{55579511401449335}{125787073112064}+\frac{44433039725 n}{108260928 j}\bigg) \eta\nonumber \\ &\qquad -\frac{240910046095 \eta^2}{518482944}\bigg) \chi^{-44/9}+\bigg(\frac{3654447011975}{98224939008}-\frac{4300262795285 \eta }{18124839936}+\frac{392328884035 \eta ^2}{1294631424}\bigg) \chi ^{-50/9} \bigg\} e_0^4,\\ \nonumber \\
&\mathcal{P}_5=\frac{14453 \pi }{756}-\frac{32 \pi n}{j}-\frac{65 \pi }{9} \eta -\bigg(\frac{1675}{756}+\frac{160 n}{3 j}+\frac{65 \eta }{9}\bigg) \pi  \log\left(\frac{f}{j}\right)+\bigg\{\bigg(-\frac{458370775 \pi }{6837264}-\frac{4909969 \pi  n}{46512 j}\nonumber \\ &\qquad+\frac{15803101 \pi  \eta }{229824}\bigg) \chi ^{-19/9}+\bigg(\frac{185734313 \pi
   }{4112640}-\frac{12915517 \pi  \eta }{146880}\bigg) \chi ^{-25/9}+\bigg(\frac{26056251325 \pi }{1077705216}+\frac{680485 \pi  n}{12384 j}\nonumber \\ &\qquad-\frac{48393605 \pi  \eta }{895104}\bigg) \chi ^{-28/9}+\bigg(-\frac{7063901 \pi }{520128}+\frac{149064749 \pi  \eta }{2210544}\bigg) \chi ^{-34/9} \bigg\}e_0^2+\bigg\{\bigg(-\frac{1523166085325 \pi }{6235584768}\nonumber \\ &\qquad-\frac{16315826987 \pi  n}{42418944 j}+\frac{52513704623 \pi  \eta }{209599488}\bigg) \chi^{-19/9}+\bigg(\frac{238457223541 \pi }{696563712}-\frac{17513506613 \pi  \eta }{33488640}\bigg) \chi ^{-25/9}\nonumber \\ &\qquad+\bigg(\frac{268377522549925 \pi
   }{1965734313984}+\frac{368891935 \pi  n}{1188864 j}-\frac{498450665645 \pi  \eta }{1632669696}\bigg) \chi ^{-28/9}+\bigg(-\frac{2408172473789 \pi }{6790791168}\nonumber \\ &\qquad+\frac{992200223893 \pi  \eta }{1697697792}\bigg) \chi ^{-34/9}+\bigg(\frac{34901256494241693175 \pi }{79386134731997184}+\frac{84423313781887 \pi  n}{193345546752 j}-\frac{15387742160333 \pi  \eta }{39404703744}\bigg) \chi ^{-38/9}\nonumber \\ &\qquad+\bigg(-\frac{17596253179825 \pi }{51451158528}+\frac{1223601085925 \pi  \eta }{1837541376}\bigg) \chi ^{-44/9}+\bigg(-\frac{7525784976509075 \pi }{38703714803712}-\frac{85031756225 \pi  n}{216521856 j}\nonumber \\ &\qquad+\frac{461030900395 \pi  \eta }{1036965888}\bigg) \chi ^{-47/9}+\bigg(\frac{14896370333 \pi
   }{61544448}-\frac{351697861441 \pi  \eta }{476969472}\bigg) \chi ^{-53/9} \bigg\}e_0^4 \\ \nonumber 
\end{align}
\end{subequations}
\end{widetext}
For the ease of presentation, we split the 3PN contributions to $\Psi_j^{n}$ 
in to three parts
%For legibility we again split the 3PN part of $\Psi_j^{+n}$ expression similar to $e_t$  as,
\begin{align}
\mathcal{P}_6= \mathcal{P}_6^{'} + \mathcal{P}_6^{''}e_0^2 + \mathcal{P}_6^{'''} e_0^4 \\ \nonumber 
\end{align} 
Various contributions to $  \mathcal{P}_6 $ are given by,    
\begin{widetext}
\begin{subequations}
\begin{align}
&\mathcal{P}_6^{'}=\frac{13966988843531}{4694215680}+\frac{257982425 n}{508032 j}-\frac{640 \pi ^2}{3}-\frac{6848 \gamma
   }{21}+\bigg(-\frac{20562265315}{3048192}-\frac{2393105 n}{1512 j}+\frac{23575 \pi ^2}{96}\nonumber \\ &\qquad+\frac{1845 \pi ^2 n}{32 j}\bigg) \eta +\bigg(\frac{110255}{1728}+\frac{475 n}{24 j}\bigg) \eta ^2-\frac{127825 \eta ^3}{1296}-\frac{13696 \log (2)}{21}-\frac{3424 \log (x)}{21},\\ \nonumber \\
&\mathcal{P}_6^{''}= \bigg\{\frac{4175723876720788380517}{5556561877278720000}+\frac{534109712725265 n}{2405438042112 j}-\frac{21508213 \pi ^2}{276480}-\frac{734341 \gamma}{16800} +\bigg(-\frac{37399145056383727}{28865256505344}\nonumber \\ &\qquad-\frac{1219797059185 n}{2045440512 j}+\frac{12111605 \pi ^2}{264192}+\frac{639805 n \pi^2}{22016 j}\bigg) \eta +\bigg(-\frac{159596464273381}{1718170030080} +\frac{43766986495 n}{1022720256 j}\bigg) \eta ^2-\frac{69237581 \eta^3}{746496}\nonumber \\ &\qquad-\frac{9663919 \log (2)}{50400}+\frac{4602177 \log (3)}{44800}-\frac{734341 \log (x)}{33600}\bigg\}\chi ^{-19/9}+\bigg\{\frac{326505451793435}{2061804036096}+\frac{916703174045 n}{5080610304 j}\nonumber \\ &\qquad-\bigg(\frac{13467050491570355}{39689727694848}+\frac{9519440485 n}{35282016 j}\bigg) \eta -\bigg(\frac{2186530635995}{52499639808} +\frac{7198355375 n}{45362592 j}\bigg) \eta ^2+\frac{2105566535 \eta^3}{10606464}\bigg\} \chi ^{-25/9}\nonumber \\ &\qquad+\frac{24716497 \pi ^2 }{293760}\chi ^{-28/9}+\bigg\{-\frac{82471214720975}{45625728024576}-\frac{2153818055 n}{524289024 j} +\bigg(-\frac{48415393035455}{1629490286592}-\frac{119702185 n}{1560384 j}\bigg) \eta \nonumber \\ &\qquad+\bigg(\frac{906325428545}{6466231296}+\frac{32769775 n}{222912 j}\bigg) \eta ^2-\frac{2330466575 \eta ^3}{16111872}\bigg\} \chi ^{-31/9} +\bigg\{-\frac{4165508390854487}{16471063977984}-\frac{96423905 \pi ^2}{5052672}\nonumber \\ &\qquad+\frac{2603845 \gamma}{61404}+\bigg(-\frac{1437364085977}{53477480448}+\frac{3121945 \pi ^2}{561408}\bigg) \eta +\frac{4499991305 \eta ^2}{636636672}+\frac{2425890995 \eta^3}{68211072}+\frac{1898287 \log (2)}{184212}\nonumber \\ &\qquad +\frac{12246471 \log (3)}{163744}+\frac{2603845 \log (x)}{122808}-\frac{2603845 \log (\chi)}{184212}\bigg\}\chi ^{-37/9}, \\ \nonumber \\
&\mathcal{P}_6^{'''}=\bigg\{\frac{13875930442343179788457991}{5067584432078192640000}+\frac{1774846575386055595 n}{2193759494406144 j}-\frac{71471791799 \pi ^2}{252149760}-\frac{2440215143 \gamma
   }{15321600}\nonumber \\ &\qquad+\bigg(-\frac{124277359022363124821}{26325113932873728}-\frac{4053385627671755 n}{1865441746944 j}+\frac{40246863415 \pi^2}{240943104}+\frac{2126072015 n \pi ^2}{20078592 j}\bigg) \eta \nonumber \\ &\qquad+\bigg(-\frac{530339050780445063}{1566971067432960}+\frac{7654615585415 n}{49090572288 j}\bigg) \eta ^2-\frac{230076481663 \eta ^3}{680804352}-\frac{32113202837 \log (2)}{45964800}\nonumber \\ &\qquad+\frac{5097678057 \log (3)}{13619200}-\frac{2440215143
   \log (x)}{30643200}\bigg\}\chi ^{-19/9}+\bigg\{\frac{2095939685244436475}{1746053475139584}+\frac{5884601777755325 n}{4302551126016 j}\nonumber \\ &\qquad+\bigg(-\frac{17381974915387486205}{8402882349109248}-\frac{527634379756765 n}{358545927168 j}\bigg) \eta +\bigg(-\frac{386694251193132845}{933653594345472}-\frac{9761006428375 n}{10342670976 j}\bigg) \eta ^2\nonumber \\ &\qquad+\frac{2855158909615 \eta ^3}{2418273792}\bigg\}\chi ^{-25/9}+\frac{254578148953 \pi ^2}{535818240} \chi ^{-28/9}+\bigg\{\frac{141251897794072110575}{3786570420215611392}+\frac{194154433667165 n}{2290094456832 j}\nonumber \\ &\qquad+\bigg(-\frac{11182467092862313645}{19319236837834752}-\frac{15348073704055 n}{13631514624 j}\bigg) \eta+\bigg(\frac{1038816664853665}{594291769344}+\frac{2534255435 n}{1741824 j}\bigg) \eta ^2\nonumber \\ &\qquad-\frac{147245442666235 \eta ^3}{102858190848}\bigg\} \chi^{-31/9}+\bigg\{\frac{102453749612934666311}{19868699733442560}-\frac{598067688595 \pi^2}{4608036864}-\frac{36290762107 \gamma }{56000448}\nonumber \\ &\qquad+\bigg(\frac{6738669506224179365}{2219101528670208}-\frac{110934582115 \pi ^2}{512004096}\bigg)\eta -\frac{1484623162301215 \eta ^2}{6604468835328}+\frac{128895671353745 \eta ^3}{217729741824}\nonumber \\ &\qquad-\frac{1140350944327 \log(2)}{24000192}+\frac{1296725746149 \log (3)}{49778176}-\frac{36290762107 \log (x)}{112000896}+\frac{36290762107 \log (\chi )}{168001344}\bigg\}\chi ^{-37/9}\nonumber \\ &\qquad+\bigg\{-\frac{3123488330286080905561719773}{355085641155718958284800}-\frac{85280660877506238107 n}{124770071244349440 j}+\frac{300051120571 \pi ^2}{970776576}+\frac{211649317 \gamma
   }{191520}\nonumber \\ &\qquad+\bigg(-\frac{40336854286157147692937}{32939298808508252160}+\frac{584462420500316711 n}{495119330334720 j}+\frac{2786391039419 \pi^2}{17972849664}-\frac{91683875075 n \pi ^2}{1089263616 j}\bigg) \eta\nonumber \\ &\qquad +\bigg(\frac{14654969487690651143}{35648591784099840}-\frac{46042929781519 n}{107385626880 j}\bigg) \eta ^2+\frac{49171400252465 \eta ^3}{91738386432}+\frac{2117998887803 \log (2)}{44241120}\nonumber \\ &\qquad-\frac{334711679031 \log
   (3)}{13108480}+\frac{211649317 \log (x)}{383040}\bigg\}\chi ^{-38/9}+\bigg\{-\frac{1017258852718193648990131}{859416250731078942720}\nonumber \\ &\qquad-\frac{284592379883138801345 n}{227358796489703424 j}+\bigg(\frac{69311096542161812013731}{30693437526109962240}+\frac{17602484074819772515 n}{12179935526234112 j}\bigg) \eta \nonumber \\ &\qquad   +\bigg(\frac{3272123415010135297}{2970715982008320}+\frac{129257754627385505 n}{66922722671616 j}\bigg) \eta ^2-\frac{40063118477671 \eta^3}{20353213440}\bigg\} \chi ^{-44/9}\nonumber \\ &\qquad-\frac{2341612230425 \pi ^2}{3675082752} \chi^{-47/9}+\bigg\{-\frac{181582918442691290125}{1374276523167055872}-\frac{157819616198875 n}{591398019072
   j}+\bigg(\frac{1741702918744309017425}{1521520436363526144}\nonumber \\ &\qquad+\frac{185709581143825 n}{109127015424 j}\bigg) \eta +\bigg(-\frac{18130335399490218365}{6037779509379072}-\frac{16942972137575 n}{7794786816 j}\bigg) \eta ^2+\frac{91862546967565 \eta^3}{37330771968}\bigg\} \chi ^{-50/9}\nonumber \\ &\qquad + \bigg\{\frac{259620437372696563}{159257838845952}+\frac{691917129965 \pi ^2}{2589262848}-\frac{558835855 \gamma}{2030112}+\bigg(-\frac{245999063921173}{13702378991616}-\frac{20770936405 \pi ^2}{575391744}\bigg) \eta\nonumber \\ &\qquad +\frac{255806950720535 \eta^2}{326247118848}-\frac{9022269087085 \eta ^3}{8738762112}-\frac{12629690323 \log (2)}{188800416}-\frac{27159422553 \log (3)}{55940864}-\frac{558835855\log (x)}{4060224}\nonumber \\ &\qquad+\frac{558835855 \log (\chi )}{6090336}\bigg\}\chi^{-56/9}. \\ \nonumber 
\end{align}
\end{subequations}
\end{widetext}
Let us emphasize that the above 
 expression indeed provides all the required Fourier phases, $\Psi_j^{\pm n}$'s that appear in 
 Eq.~(\ref{hf_1PN}) for 
 $\tilde{h}(f)$. For instance, Fourier phases present in the quadrupolar order $\tilde{h}(f)$, namely, $\Psi_j^{0}$, $\Psi_j^{+2}$ and $\Psi_j^{-2}$ are  obtained by putting in Eq.~(\ref{appendixpsi}) 
 $n=0,+2,-2$, respectively. 
 Further, 
 one should evaluate these Fourier phases at the correct stationary points and this requires us to use  $x=\bigg\{\frac{G\,m\,2\,\pi\,f}{c^3\,\left(j-(j \pm n)k^{(6)}_{(3)}\right)}\bigg\}^{2/3}$.
 We note in passing that 
 the 3PN accurate $e_t$ and $\Psi_j^{n}$ expressions
 along with 1PN accurate Fourier amplitudes while incorporating 
 eccentricity corrections to $\mathcal{O}(e_0^6)$ at each PN order can be found in the attached \texttt{Mathematica} notebook.

%\nocite{*}
%\bibliographystyle{apsrev4-1}
\bibliography{TGMH_v2}
\end{document}